\documentclass[12pt]{article}
\usepackage{amsmath,amssymb}
\usepackage[mathscr]{euscript}

\usepackage[dvips,final]{graphicx}
\usepackage[dvips]{geometry}
\usepackage{color}
\usepackage{epsfig}
\usepackage{latexsym}
\usepackage{pstricks}
\newtheorem{defi}{Definition}[section]

\newcommand{\iN}{\hbox{ {\leaders\hrule\hskip.2cm}{\vrule height
      .22cm} }}

\newcommand{\R}{\mathbb{R}}
\newcommand{\C}{\mathbb{C}}

\newcommand{\gP}{\mathfrak{P}}
\newcommand{\gsp}{\mathfrak{sp}}

\title{Multisymplectic formalism and the covariant phase space}
\author{Fr{\'e}d{\'e}ric \textsc{H{\'e}lein}\footnote{Institut de Math{\'e}matiques de Jussieu, UMR CNRS 7586 Universit{\'e} Denis Diderot Paris 7,
175 rue du Chevaleret, 75013 Paris, France, \textsf{helein@math.jussieu.fr}}
}

\begin{document}
\maketitle

In most attempts for building the mathematical foundations of Quantum Fields Theory (QFT) two classical ways have been explored. The first one is often referred to as the \emph{Feynman integral} or \emph{functional integral method}. It is a generalization to fields of the path integral method for quantum mechanics and is heuristically based on computing integrals over the infinite dimensional set of all possible fields $\phi$ by using a kind of `measure' --- which should behave like the Lebesgue measure on the set of all possible fields $\phi$ --- times $e^{i\mathcal{L}(\phi)/\hbar}$, where $\mathcal{L}$ is a Lagrangian functional (but attempts to define this `measure' failed in most cases). The second one is referred to as the \emph{canonical quantization method} and is based on the Hamiltonian formulation of the dynamics of classical fields, by following general axioms which were first proposed by Dirac and later refined. The Feynman approach has the advantage of being manifestly relativistic, i.e. it does not require the choice of a particular system of space-time coordinate, since the main ingredient is $\mathcal{L}(\phi)$, which is an integral over all space-time. However the canonical approach, at least its classical formulation, seems to be based to the choice of a particular time coordinate which is needed to define the Hamiltonian function through an infinite dimensional Legendre transform.

However they are alternative formulations of the Hamilton structure of the dynamics of classical fields, which could be used as a starting point of a \emph{covariant canonical quantization}\footnote{The word `\emph{covariant}' refers here to a construction which does rely on the choice of a particular system of coordinates on space-time and hence which respects the basic principles of Relativity.}.  We shall see two of them in this text: the \emph{covariant phase space} and the \emph{multisymplectic formalism}. The covariant phase space is based on the observation that the set of classical solutions to a variational problem, i.e. of critical points of some action functional $\mathcal{L}(\phi)$, has an intrinsic canonical symplectic (or presymplectic) structure. The multisymplectic formalism is a generalization of the standard symplectic formalism, where the \emph{time} in classical mechanics is replaced by the \emph{space-time}: for instance if we start from a Lagrangian action $\mathcal{L}(\phi) = \int_\mathcal{X} L(x,\phi,d\phi)$ we do not perform a Legendre transform with respect to a chosen time coordinates, but with respect to all space-time coordinates.

We expect that roughly speaking both the Feynman and the canonical approach should lead to equivalent theories. However if this fact is true, it should not be trivial for several reasons. A first obvious remark for that is that both theories are only heuristics and have no mathematical foundations, excepted in very simplified situations. A superficial difference between both approaches is the fact that one is based on the Lagrangian, the other one the Hamiltonian function. Moreover these two approaches answer to different questions, the Feynman offers a short and intuitive way to compute the quantities which can be measured in interaction processes between particles (although one cannot avoid the difficult step of regularizing and renormalizing the computed quantities). For the same task, the canonical approach seems to be more complicated, however it proposes a scheme to build mathematical objects (a complex Hilbert space of physical \emph{states} and an algebra of self-adjoint operators acting on it, corresponding to \emph{observable quantities}), the construction of which requires more effort by using the Feynman integral. But a deep difference between both methods is that the Feynman integral is a construction \emph{off shell}, i.e. on the set of \emph{all} possible fields, even those which are not solutions of the classical dynamical equations, whereas in many cases the canonical approach is a construction \emph{on shell}, i.e. on the set of fields which are solutions of the dynamical equations (in particular in the covariant phase space method).

In this paper we shall present briefly the multisymplectic formalism and the covariant phase space and show the strong relation between both theories. To my knowledge this relation was discovered by J. Kijowski and W. Szczyrba in 1976 \cite{KijowskiSzczyrba}, but their beautiful paper seems to have been ignored in the literature. We have included some historical comments. We shall conclude by presenting the geometric quantization scheme for linear field equations (i.e. \emph{free fields} in the terminology of physicists) in the framework of multisymplectic geometry. The goal is to show how a canonical quantization could be performed in covariant way.

\section{The multisymplectic formalism}
\subsection{Maps between vector spaces}\label{paragraph1.1}
We start with a simple variational problem: let $\textsc{X}$ and $\textsc{Y}$ be two vector space of dimension $n$ and $k$ respectively and assume that $\textsc{X}$ is oriented, let $U$ be an open subset of $\textsc{X}$ and consider the set $\mathcal{C}^\infty(U,\textsc{Y})$ of smooth maps $\mathbf{u}$ from $U$ to $\textsc{Y}$. Let $L:U\times \textsc{X}\times \hbox{End}(\textsc{X},\textsc{Y})\longrightarrow \mathbb{R}$ be a Lagrangian density and consider the action functional on $\mathcal{C}^\infty(U,\textsc{Y})$ defined by:
\[
\mathcal{L}[\mathbf{u}] = \int_UL(x,\mathbf{u}(x),d\mathbf{u}_x)\beta,
\]
where $\beta$ is a volume $n$-form on $U$. We use coordinates $(x^1,\cdots,x^n)$ on $U$ s.t. $\beta = dx^1\wedge \cdots \wedge dx^n$, coordinates $(y^1,\cdots,y^k)$ on $\textsc{Y}$ and $v^i_\mu$ on $\hbox{End}(\textsc{X},\textsc{Y})$. Then the critical points of $\mathcal{L}$ satisfy the Euler--Lagrange system of equations
\begin{equation}\label{euler-lagrange}
\frac{\partial}{\partial x^\mu}\left(\frac{\partial L}{\partial v^i_\mu}(x,\mathbf{u}(x),d\mathbf{u}_x)\right) = \frac{\partial L}{\partial y^i}(x,\mathbf{u}(x),d\mathbf{u}_x),\quad \forall i\hbox{ s.t. }1\leq i\leq k.
\end{equation}
We assume that the map
\[
\begin{array}{ccl}
U\times \textsc{Y}\times \hbox{End}(\textsc{X},\textsc{Y}) & \longrightarrow & U\times \textsc{Y}\times \hbox{End}(\textsc{Y}^*,\textsc{X}^*)\\
(x,y,v) & \longmapsto & (x,y,\frac{\partial L}{\partial v}(x,y,v)),
\end{array}
\]
is a diffeomorphism: this is the analogue of the \emph{Legendre hypothesis} used in Mechanics. We denote by $p^* = (p^\mu_i)_{\mu,i}$ the coordinates on $\hbox{End}(\textsc{Y}^*,\textsc{X}^*)$ and we define the Hamiltonian function $H$ on $U\times \textsc{X}\times \hbox{End}(\textsc{Y}^*,\textsc{X}^*)$ by
\[
H(x,y,p^*):= p^\mu_iv^i_\mu - L(x,y,v),
\]
where we assume implicitely that $v = (v^i_\mu)_{i,\mu}$ is the unique solution of $\frac{\partial L}{\partial v}(x,y,v) = p^*$. Then $v^i_\mu$ is actually equal to $\frac{\partial H}{\partial p^\mu_i}$. Moreover to any map $\mathbf{u}:U\longrightarrow \textsc{Y}$ we associate the map $\mathbf{p}^*:U\longrightarrow \hbox{End}(\textsc{Y}^*,\textsc{X}^*)$ s.t. $\mathbf{p}^*(x):= \frac{\partial L}{\partial v}(x,\mathbf{u}(x),d\mathbf{u}_x)$, $\forall x\in U$. Then we can show \cite{volterra1} that $\mathbf{u}$ is a solution of (\ref{euler-lagrange}) iff $(\mathbf{u},\mathbf{p}^*)$ is a solution of the generalized Hamilton system:
\begin{equation}\label{Hamilton0}
\left\{
\begin{array}{ccl}
\displaystyle \frac{\partial \mathbf{u}^i}{\partial x^\mu}(x) & = & \displaystyle \frac{\partial H}{\partial p^\mu_i}(x,\mathbf{u}(x),\mathbf{p}^*(x))\\
\displaystyle \frac{\partial \mathbf{p}^\mu_i}{\partial x^\mu}(x) & = & \displaystyle - \frac{\partial H}{\partial y^i}(x,\mathbf{u}(x),\mathbf{p}^*(x)).
\end{array}\right.
\end{equation}
System (\ref{Hamilton0}) can be translated as a geometric condition \cite{Kijowski1} on the graph
\[
\Gamma^*:= G(\mathbf{u},\mathbf{p}^*):= \{(x,\mathbf{u}(x),\mathbf{p}^*(x))|\ x\in U\}\subset U\times \textsc{Y}\times \hbox{End}(\textsc{Y}^*,\textsc{X}^*).
\]
Indeed consider a family of $n$ vector fields $X_1,\cdots ,X_n:U\longrightarrow U\times \textsc{Y}\times \hbox{End}(\textsc{Y}^*,\textsc{X}^*)$ s.t. for any $x\in U$, $(X_1(x),\cdots ,X_n(x))$ is a basis of the tangent plane to $G(\mathbf{u},\mathbf{p}^*)$ at $(x,\mathbf{u}(x),\mathbf{p}^*(x))$. Set $\beta_\mu:= \frac{\partial}{\partial x^\mu}\iN \beta$. Then (\ref{Hamilton0}) is equivalent to the condition that $\forall \xi\in \textsc{X}\times \textsc{Y}\times \hbox{End}(\textsc{Y}^*,\textsc{X}^*)$,
\begin{equation}\label{Hamilton1}
dp^\mu_i\wedge dy^i\wedge \beta_\mu(\xi,X_1,\cdots,X_n) = dH(\xi)\beta(X_1,\cdots,X_n).
\end{equation}
In fact this can be easily checked by choosing $X_\mu:= \frac{\partial}{\partial x^\mu} + \frac{\partial \mathbf{u}^i}{\partial x^\mu}\frac{\partial}{\partial y^i} + \frac{\partial \mathbf{p}^\mu_i}{\partial x^\mu}\frac{\partial}{\partial p^\mu_i}$. More concisely we can introduce the $n$-multivector field $X:= X_1\wedge \cdots \wedge X_n$ (so that $X(x)\in \Lambda^nT_{(x,\mathbf{u}(x),\mathbf{p}^*(x))}\Gamma^*$, $\forall x\in U$). Then Equation (\ref{Hamilton1}) reads:
\begin{equation}\label{Hamilton1bis}
\forall \xi\in \textsc{X}\times \textsc{Y}\times \hbox{End}(\textsc{Y}^*,\textsc{X}^*),\quad
dp^\mu_i\wedge dy^i\wedge \beta_\mu(\xi,X) = dH(\xi)\beta(X).
\end{equation}
Equation (\ref{Hamilton1}) can be completed with the independence condition
\begin{equation}\label{independence}
\beta|_{\Gamma^*} \neq 0,
\end{equation}
where, if $j_{\Gamma^*}:\Gamma^*\longrightarrow U\times \textsc{Y}\times \hbox{End}(\textsc{Y}^*,\textsc{X}^*)$ denotes the inclusion map, $\beta|_{\Gamma^*}:= j_{\Gamma^*}^*\beta$. This condition garantees that locally $\Gamma^*$ is the graph of some map $(\mathbf{u},\mathbf{p}^*)$ over the `space-time' $\textsc{X}$.

We will see now that the independence condition (\ref{independence}) can be further incorporated in a dynamical condition analogous to (\ref{Hamilton1}) by adding to the variables $(x,y,p^*)$ a variable $e$ dual to $\beta$. We define $\textsc{M}:= U\times \textsc{Y}\times \mathbb{R} \times \hbox{End}(\textsc{Y}^*,\textsc{X}^*)$ with coordinates $(x,y,e,p^*) = (x^\mu,y^i,e,p^\mu_i)$ and the $(n+1)$-form
\begin{equation}\label{n+1formedebase}
\omega:= de\wedge \beta + dp^\mu_i\wedge dy^i\wedge \beta_\mu.
\end{equation}
We define $\mathcal{H}:\textsc{M}\longrightarrow \mathbb{R}$ by $\mathcal{H}(x,y,e,p^*):= e + H(x,y,p^*)$. Then to any oriented $n$-dimensional submanifold $\Gamma^* = G(\mathbf{u},\mathbf{p}^*)$ we associate the oriented $n$-dimensional submanifold $\Gamma:= \{(x,\mathbf{u}(x),\mathbf{e}(x),\mathbf{p}^*(x))|\ x\in U\}$ of $\textsc{M}$, where $\mathbf{e}$ is s.t. $\mathbf{e}(x) + H(x,\mathbf{u}(x),\mathbf{p}^*(x))= h$, $\forall x\in \textsc{X}$, for some real constant\footnote{W.l.g. we can assume that the constant $h$ is zero, so that $\Gamma$ is included in $\mathcal{H}^{-1}(0)$.} $h$. Then $\Gamma^*$ is a solution of (\ref{Hamilton1}) and (\ref{independence}) iff $\Gamma$ is a solution of:
\begin{equation}\label{Hamilton2}
\forall \xi\in \mathcal{C}^\infty(\textsc{M},T\textsc{M}),\quad
\omega(\xi,X) = d\mathcal{H}(\xi)\beta(X),
\end{equation}
where $\mathcal{C}^\infty(\textsc{M},T\textsc{M})$ denotes the set of sections of $T\textsc{M}$ over $\textsc{M}$, which can be identified with $\mathcal{C}^\infty(\textsc{M},\textsc{X}\times \textsc{Y}\times \mathbb{R}\times \hbox{End}(\textsc{Y}^*,\textsc{X}^*))$.
Note that conversely it is easy to check that any connected solution $\Gamma$ of (\ref{Hamilton2}) is contained in a level set of $\mathcal{H}$. We call a \textbf{Hamiltonian $n$-curve} any solution $\Gamma$ of (\ref{Hamilton2}). The $(n+1)$-form $\omega$ is an example of a multisymplectic form and the pair $(\textsc{M},\omega)$ is called a multisymplectic manifold. Using the notation $\xi\iN \omega$ for the interior product of the vector $\xi$ with the $(n+1)$-form $\omega$, we set:
\begin{defi}
Let $\mathcal{M}$ be a smooth manifold. A \textbf{multisymplectic $(n+1)$-form} $\omega$ on $\mathcal{M}$ is a $(n+1)$-form which is \textbf{closed} (i.e. $d\omega = 0$) and which is \textbf{non degenerate} (i.e. $\forall \textsc{m}\in \mathcal{M}$, $\forall \xi\in T_\textsc{m}\mathcal{M}$, $\xi\iN \omega = 0 \Longrightarrow \xi =0$).
\end{defi}

\subsection{Higher order problems}\label{paragraph1.2}
The preceding can be extended to an action on maps $\mathbf{u}:U\longrightarrow \textsc{Y}$ of the form $\mathcal{L}[\mathbf{u}]:= \int_UL(x,j^r\mathbf{u}(x))\beta$, where $j^r\mathbf{u}$ denotes the $r$-th order jet of $\mathbf{u}$ (i.e. all partial derivatives of $\mathbf{u}$ of order less than or equal to $r$). We denote by $J^r(U,\textsc{Y})$ the $r$-th order jet space of maps from $U$ to $\textsc{Y}$ and we use the coordinates $x = (x^\mu)_\mu$ and  $v = \left(v^i_{\mu_1\cdots \mu_a}\right)_{i,\mu_1\cdots \mu_a}$ (for $1\leq i\leq k$, $0\leq a\leq r$ and $1\leq \mu_b\leq n$) on $J^r(U,\textsc{Y})$ s.t.
\[
v^i_{\mu_1\cdots \mu_a}\left(j^r\mathbf{u}(x)\right) = \frac{\partial^a \mathbf{u}^i}{\partial x^{\mu_1}\cdots \partial x^{\mu_a}}(x).
\]
It is convenient to introduce the multi-index notation $M = \mu_1\cdots \mu_a$, where $a\in \mathbb{N}$ and $\forall b\in [\![ 1,a]\!]$, $1\leq \mu_b \leq n$ and to set $|M| = a$. Then for $|M|= r$ we note
\begin{equation}\label{Legendreordrer}
\pi^M_i(x,v):= \frac{\partial L}{\partial v^i_M}(x,v).
\end{equation}
The analogue of the Legendre hypothesis consists here in supposing that the map $(v^i_M)_{i,M;|M|=r}\longmapsto (p^M_i)_{i,M;|M|=r}$ defined by (\ref{Legendreordrer}) is one to one. Next we define the vector space $\textsc{M}$ with coordinates:
\[
\begin{array}{ccccccccc}
(x,v) & : &x^\mu&  & v^i & v^i_\mu & \cdots & v^i_{\mu_1\cdots \mu_{r-1}} & \\
p = (e,p^*) & : & & e &  & p^\mu_i & \cdots & p_i^{\mu_1\cdots \mu_{r-1}} & p_i^{\mu_1\cdots \mu_r} 
\end{array}
\] 
for $1\leq \mu,\mu_b\leq n$ and $1\leq i\leq k$. Clearly $\textsc{M}$ contains $J^{r-1}(U,\textsc{Y})$ as a vector subspace. We also define recursively, for $|M|\leq r-1$,
\[
\pi^M_i(x,v):= \frac{\partial L}{\partial v^i_M}(x,v)
- D_\mu\pi^{M\mu}_i(x,v_M),
\]
where
\[
D_\mu:= \frac{\partial}{\partial x^\mu} + v^i_{M\mu}\frac{\partial}{\partial v^i _M}.
\]
Then we define a Hamiltonian function on $\textsc{M}$:
\[
H(x,v,p^*):= p^\mu_iv^i_\mu + \cdots + p_i^{\mu_1\cdots \mu_r}v^i_{\mu_1\cdots \mu_r} - L(x,v),
\]
where we assume implicitely that, for $|M|=r$, $v^i_M = v^i_{\mu_1\cdots \mu_r}$ is the solution of $p_i^M = \pi^M_i(x,v)$, $\forall M$ s.t. $|M| = r$, and we set $p^M_i:= \pi^M_i(x,v)$, $\forall M$ s.t. $|M|<r$.
To any map $\mathbf{u}$ from $U$ to $\textsc{Y}$ we associate the map $\mathbf{p}^*$ which is the image of $j^r\mathbf{u}$ by the maps $\pi^M_i$. Then $\mathbf{u}$ is a critical point of $\mathcal{L}$ iff $(j^{r-1}\mathbf{u},\mathbf{p}^*)$ is a solution of the generalized Hamilton equations \cite{DeDonder}
\begin{equation}\label{HamiltonToutOrdre}
\left\{
\begin{array}{ccll}
\displaystyle \frac{\partial \mathbf{u}^i_{\mu_1\cdots\mu_a}}{\partial x^\mu}(x) & = & \displaystyle \frac{\partial H}{\partial p^{\mu_1\cdots\mu_a\mu}_i}(x,\mathbf{u}(x),\mathbf{p}^*(x)) & \hbox{for }0\leq a \leq r-1\\
\displaystyle \frac{\partial \mathbf{p}^{\mu_1\cdots\mu_a\mu}_i}{\partial x^\mu}(x) & = & \displaystyle - \frac{\partial H}{\partial v^i_{\mu_1\cdots\mu_a}}(x,\mathbf{u}(x),\mathbf{p}^*(x)) & \hbox{for }0\leq a \leq r-1,
\end{array}\right.
\end{equation}
Alternatively we can consider the map $x\longmapsto (x,j^{r-1}\mathbf{u}(x),\mathbf{e}(x),\mathbf{p}^*(x))$, where $\mathbf{e}$ may be chosen so that $\mathbf{e}(x) + H(x,j^{r-1}\mathbf{u}(x),\mathbf{p}^*(x)) = 0$, $\forall x$ and we can write (\ref{HamiltonToutOrdre}) in a way similar to (\ref{Hamilton2}) by using the Hamiltonian function
\[
\mathcal{H}(x,v,p) = e + H(x,v,p^*).
\]
and the multisymplectic form
\[
\omega:= de\wedge \beta + dp^\mu_i\wedge dv^i\wedge \beta_\mu + \cdots + dp_i^{\mu_1\cdots \mu_r}\wedge dv^i_{\mu_1\cdots \mu_{r-1}}\wedge \beta_{\mu_r}.
\]
An intrinsic geometrical multisymplectic formulation of these equations has been derived recently by L. Vitagliano \cite{vita1}.

\subsection{More general multisymplectic manifolds}\label{paragraph1.3}

Assume that we start with an action $\mathcal{L}$ which is an integral of a Lagrangian density which depends on the first order derivatives of the field. This may be for instance a variational problem on maps $\mathbf{u}$ between two manifolds $\mathcal{X}$ and $\mathcal{Y}$ with a functional $\mathcal{L}[\mathbf{u}]:= \int_\mathcal{X} L(x,\mathbf{u}(x),d\mathbf{u}_x)\beta$ or a variational problem on sections of a fiber bundle $\pi: \mathcal{Z}\longrightarrow \mathcal{X}$. Then a natural multisymplectic manifold is the vector bundle  $\Lambda^{n}T^*(\mathcal{X}\times \mathcal{Y})$ in the first case or $\Lambda^nT^*\mathcal{Z}$ in the second case. Both manifolds are indeed endowed with a canonical $(n+1)$-form $\omega$ which is the straightforward analogue of the canonical symplectic form on any cotangent bundle \cite{Dedecker,Kijowski2,hk1}. We may call this manifold the \emph{universal multisymplectic manifold} associated with the Lagrangian problem. Although this construction seems to be similar to the symplectic one for Hamiltonian mechanics, it is different because, say for maps between two manifolds $\mathcal{X}$ and $\mathcal{Y}$ of dimensions $n$ and $k$ respectively and a first order variational problem, on the one hand the Lagrangian density depends on $n+k+nk$ variables (in other words the analogue of the product of the time real line and of the tangent bundle in mechanics has dimension $n+k+nk$), whereas on the other hand the analogue of the cotangent bundle is $\Lambda^nT^*(\mathcal{X}\times \mathcal{Y})$ and has dimension $n+k+ \frac{(n+k)!}{n!k!}$. This means that we have much more choices in the Legendre transform, which is not a map in general but a correspondence, as soon as $n\geq 2$ and $k\geq 2$.

This is why it is possible and often simpler to impose extra constraints in the Legendre transform, which means that we replace the universal multisymplectic manifold $\Lambda^nT^*\mathcal{Z}$ (whatever $\mathcal{Z}$ is: a Cartesian product $\mathcal{X}\times \mathcal{Y}$ or the total space of a bundle) by some submanifold of it. Most Authors prefer to use the \emph{affine multisymplectic submanifold} $\Lambda^n_2T^*\mathcal{Z}$: if $\mathcal{Z} = \mathcal{X}\times \mathcal{Y}$, it is the subbundle of $\Lambda^{n}T^*(\mathcal{X}\times \mathcal{Y})$ over $\mathcal{X}\times \mathcal{Y}$, the fiber over the point $(x,y)\in \mathcal{X}\times \mathcal{Y}$ of which is the subspace of $n$-forms $p\in \Lambda^nT^*_{(x,y)}(\mathcal{X}\times \mathcal{Y})$ s.t. $\forall \eta_1,\eta_2\in T_y\mathcal{Y}$, $(0,\eta_1)\wedge (0,\eta_2)\iN p =0$. It $\mathcal{Z}$ is a fiber bundle over $\mathcal{X}$, $\Lambda^n_2T^*\mathcal{Z}$, which is the subbundle over $\mathcal{Z}$, the fiber over $z\in \mathcal{Z}$ of which is the subspace of $n$-forms $p\in \Lambda^nT^*_z\mathcal{Z}$ s.t. for any pair of \emph{vertical} vectors $\eta_1,\eta_2\in T_z\mathcal{Z}$, $\eta_1\wedge \eta_2\iN p = 0$ (by `vertical' we mean that $\eta_1$ and $\eta_2$ are maps to $0\in T_{\pi(z)}\mathcal{X}$ by the differential of $\pi$). In both cases an $n$-form reads $p = e\beta + p^\mu_idy^i\wedge \beta_\mu$ in local coordinates and the latter theory is actually the right generalization of (\ref{n+1formedebase}). This theory is usually refered to as the \emph{De Donder--Weyl theory} although it was discovered by V. Volterra (see \S \ref{paragraph1.8}). Note that $\Lambda^n_2T^*\mathcal{Z}$ can alternatively be defined as being the affine dual of the first jet bundle of sections of $\mathcal{Z}$ over $\mathcal{X}$ \cite{GIMMSY}.

\subsection{Premultisymplectic manifolds}\label{paragraph1.4}
A variant consists in manifolds equipped with a closed $(n+1)$-form but without assuming a non-degeneracy condition, as for instance in \cite{KijowskiSzczyrba}:
\begin{defi}
We call a triple $(\mathcal{M},\omega,\beta)$ an \textbf{$n$-phase space} if $\mathcal{M}$ is a manifold, $\omega$ is a closed $(n+1)$-form, called a \textbf{premultisymplectic form} and $\beta$ is a non vanishing $n$-form.
\end{defi}
Examples of premultisymplectic manifolds can be built easily by starting from a multisymplectic manifold $(\mathcal{M},\omega)$ with a Hamiltonian function $\mathcal{H}$ on it which has no critical points (as for instance $\mathcal{H}(x,y,e,p^*) = e + H(x,y,p^*)$ for the previous theory). Then we let $\eta$ be a vector field on $\mathcal{M}$ s.t. $d\mathcal{H}(\eta) = 1$ everywhere and we set $\beta:= \eta\iN \omega$. For any $h\in \mathbb{R}$ the level set $\mathcal{M}^h:= \mathcal{H}^{-1}(h)$ is a submanifold. Then  $(\mathcal{M}^h,\omega|_{\mathcal{M}^h},\beta|_{\mathcal{M}^h})$ is a premultisymplectic manifold \cite{heleinprep}. In particular $\omega|_{\mathcal{M}^h}$ is obsviously closed but may be degenerate in general: indeed if $\Gamma$ is a Hamiltonian $n$-curve contained in $\mathcal{M}^h$ then any vector tangent to $\Gamma$ is in the kernel of $\xi\longmapsto \xi\iN \omega$. In fact an $n$-phase space $(\mathcal{M},\omega)$ carries an intrinsic dynamical structure: we say that \emph{an $n$-dimensional submanifold $\Gamma$ of $\mathcal{M}$ is a Hamiltonian $n$-curve if:}
\begin{equation}\label{Hamilton3}
\forall v\in \mathcal{C}^\infty(\mathcal{M},T_\textsc{m}\mathcal{M}),\quad
\left(v\iN \omega\right)|_\Gamma = 0\quad \hbox{and }\beta|_{\Gamma}\neq 0.
\end{equation}
This definition is motivated by the fact that (if $\Gamma$ is connected) $\Gamma$ is a solution of (\ref{Hamilton2}) iff there exists some $h\in \mathbb{R}$ s.t. $\Gamma$ is contained in $\mathcal{M}^h$ and $\Gamma$ is a Hamiltonian $n$-curve in the $n$-phase space $(\mathcal{M}^h,\omega|_{\mathcal{M}^h},\beta|_{\mathcal{M}^h})$ (see \cite{heleinprep}). However there are examples of premultisymplectic manifolds which do no arise from this construction as for instance the example in \cite{hk2,Rovelli} obtained by starting from the Palatini formulation of gravity.

\subsection{Action principle}\label{paragraph1.5}

We assume here that we are working in an premultisymplectic manifold $(\mathcal{M},\omega,\beta)$ and that the form $\omega$ is exact, i.e. is of the form $\omega = d\theta$, where $\theta$ is an $n$-form on $\mathcal{M}$. This is true e.g. in a submanifold of $\Lambda^nT^*\mathcal{Z}$, where $\omega$ is precisely defined as the differential of a canonical `Poincar{\'e}--Cartan' form $\theta$. To any oriented $n$-dimensional submanifold $\Gamma$ we associate the action
\begin{equation}\label{action}
\mathcal{A}[\Gamma]:= \int_\Gamma\theta.
\end{equation}
One can then show that any $n$-dimensional submanifold $\Gamma$ on which $\beta$ does not vanish is a critical point of $\mathcal{A}$ iff it is a Hamiltonian $n$-curve, i.e. a solution of (\ref{Hamilton3}) (see \cite{heleinprep}). Actually if $\Gamma$ is the image of a given configuration by some Legendre transform, then $\mathcal{A}[\Gamma]$ coincides with the Lagrangian action of the configuration we started with \cite{hk1}. Note that in the case where $\omega$ is not exact one could define a similar action on a homology class of $n$-dimensional submanifolds by replacing $\int_\Gamma$ by $\int_\Delta\omega$, where $\Delta$ is a $(n+1)$-chain connecting $\Gamma$ with a particular $n$-dimensional submanifold which generates the homology class.

\subsection{Observable functionals}\label{paragraph1.6}
An observable functional is a functional on the `space' of all solutions: this notion will be central in the next section concerning the covariant phase space. A particular class of such functionals arise in the context of multisymplectic manifolds or premultisymplectic manifolds as follows. In the following we denote by $\mathcal{F}$ the set of $n$-dimensional oriented submanifolds (\emph{fields}) and $\mathcal{M}$ and by $\mathcal{E}$ the subset of $\mathcal{F}$ composed of Hamiltonian $n$-curves.\\

\noindent
\textbf{In a multisymplectic manifold $(\mathcal{M},\omega)$}\\

We define an \textbf{infinitesimal symplectomorphism of $(\mathcal{M},\omega)$} to be a vector field $\xi\in \mathcal{C}^\infty(\mathcal{M},T\mathcal{M})$ s.t. $L_\xi\omega = 0$ (i.e. the Lie derivative of $\omega$ by $\xi$ vanishes). Note that since $\omega$ is closed, this relation is equivalent to $d(\xi\iN \omega) =0$. An important case occurs when $\xi\iN\omega$ is exact: then there exists an $(n-1)$-form $F$ 
s.t.
\begin{equation}\label{defdeF}
dF + \xi\iN \omega =0.
\end{equation}
Any $(n-1)$-form $F$ on $\mathcal{M}$ s.t. there exists a vector field $\xi$ satisfying (\ref{defdeF}) is called an \textbf{observable $(n-1)$-form}. In the case where $n=1$ then $F$ is a function and in fact any function on $\mathcal{M}$ is an observable 0-form because the symplectic form is non degenerate. However if $n\geq 2$ then an arbitrary $(n-1)$-form on $\mathcal{M}$ is \emph{not observable in general}, but if it is so then the vector field $\xi$ s.t. (\ref{defdeF}) holds is \emph{unique}: we shall denote it by $\xi_F$. Observable $(n-1)$-forms can be integrated over hypersurfaces in an $n$-curve to produce observable functionals. For that purpose, given some Hamiltonian function $\mathcal{H}$ on $\mathcal{M}$ we define a \emph{slice} $\Sigma$ to be codimension one submanifold of $\mathcal{M}$ s.t. for any Hamiltonian $n$-curve $\Gamma$ the intersection of $\Sigma$ with $\Gamma$ is transverse. We also assume that $\Sigma$ is co-oriented, which means that $\forall \textsc{m}\in \Sigma$ the 1-dimensional quotient space $T_\textsc{m}\mathcal{M}/T_\textsc{m}\Gamma$ is oriented. Then we can endow $\Sigma\cap\Gamma$ with an orientation and define
\[
\begin{array}{cccl}
\displaystyle \int_\Sigma F: & \mathcal{F} & \longrightarrow & \mathbb{R}\\
& \Gamma & \longmapsto & \displaystyle \int_{\Sigma\cap \Gamma}F
\end{array}
\]
Then one can recover two important notions in the semi-classical theory of fields. First one can define a bracket between observable $(n-1)$-forms $F$ and $G$ by the formula
\[
\{F,G\}:= \xi_F\wedge \xi_G\iN \omega = \xi_F\iN dG = - \xi_G\iN dF.
\]
Obviously $\{F,G\}$ is also an $(n-1)$-form. Moreover one can prove that it is also observable and that $\xi_{\{F,G\}} = [\xi_F,\xi_G]$ \cite{Kijowski1,hk2}. Then the set of observable $(n-1)$-forms equipped with this `Poisson bracket' becomes almost a Lie algebra (it satisfies the antisymmetry relation $\{F,G\} + \{G,F\} = 0$, but not the Jacobi identity; we have instead $\{\{G,H\},F\} + \{\{H,F\},G\} + \{\{F,G\},H\} = d(\xi_F\wedge \xi_G\wedge \xi_H\iN \omega)$, which, in the case where $n=2$ can be understood as a Lie 2-algebra structure \cite{Baez}). However we can define the bracket
\begin{equation}\label{def-poisson-multi}
\left\{\int_\Sigma F,\int_\Sigma G\right\}:= \int_\Sigma \{F,G\}
\end{equation}
which coincides with the Poisson bracket on functionals on fields used by physicists. We will also meet an interpretation of this bracket in the next Section.

A second important notion is the relation between observable forms and the dynamics. Indeed if $\Gamma$ is a Hamiltonian $n$-curve and if $F$ is an observable $(n-1)$-form then one can use the dynamical equation (\ref{Hamilton2}) with the vector field $\xi_F$. It gives us, $\forall \textsc{m}\in \Gamma$, $\forall X\in \Lambda^nT_\textsc{m}\Gamma$,
\begin{equation}\label{dFetdynamics}
dF(X) = - \omega(\xi_F,X) = - d\mathcal{H}(\xi_F)\beta(X).
\end{equation}
Hence we see that \emph{if $d\mathcal{H}(\xi_F)$ vanishes}, then $dF|_\Gamma$ vanishes. This implies by using Stokes theorem that the restriction of the functional $\int_\Sigma F$ to the set $\mathcal{E}$ of Hamiltonian $n$-curves does not depend on $\Sigma$ but on its homology class. For that reason we say that \textbf{an observable $(n-1)$-form $F$ is dynamical if $d\mathcal{H}(\xi_F) = 0$}.\\

\noindent
\textbf{In a premultisymplectic manifold $(\mathcal{M},\omega,\beta)$}\\

The definition of an observable $(n-1)$-form $F$, of the bracket and of the observable functionals $\int_\Sigma F$ can be adapted \emph{mutatis mutandis} to the case of an premultisymplectic manifold $(\mathcal{M},\omega,\beta)$. The difference is that in such a space the dynamical condition $d\mathcal{H}(\xi) = 0$ is empty (think that $\mathcal{M}$ is the level set of some Hamiltonian function $\mathcal{H}$ on a multisymplectic manifold, then the fact that $\xi$ is tangent to this level set forces it to be in the kernel of $d\mathcal{H}$). Hence \emph{any observable $(n-1)$-form is a dynamical one}.

Moreover if $\omega$ is exact, i.e. $\omega = d\theta$, we know that Hamiltonian $n$-curve are critical points of the action (\ref{action}). We can thus see that dynamical observable $(n-1)$-forms correspond to symmetries of the variational problems and the conservation law $dF|_\Gamma = 0$ for a Hamiltonian $n$-curve $\Gamma$ is nothing but Noether's first theorem \cite{Noether,yks}. Indeed  for any observable $(n-1)$-form $F$,
\[
L_{\xi_F}\theta = d(\xi_F\iN\theta) + \xi_F\iN d\theta = d(\xi_F\iN\theta) + \xi_F\iN \omega = d(\xi_F\iN\theta - F).
\]
Hence $L_{\xi_F}\theta$ is exact, so that $\xi_F$ is a symmetry of the action $\int_\Gamma\theta$ up to a divergence term. The conserved current is just $F|_\Gamma$.

\subsection{Hamilton--Jacobi equations}\label{paragraph1.7}
The Hamilton--Jacobi equation for a Hamiltonian function $\mathcal{H}$ on a multisymplectic manifold of the form $\Lambda^nT^ *\mathcal{Z}$ (or on a submanifold of it) is the following equation on an $(n-1)$-form $S$ on $\mathcal{Z}$ (i.e. a section of the vector bundle $\Lambda^{n-1}T^* \mathcal{Z}\longrightarrow \mathcal{Z}$):
\begin{equation}\label{HamiltonJacobi}
\mathcal{H}(z,dS_z) = 0.
\end{equation}
Alternatively the unknown may be chosen to be $\lambda:= dS$: we then require that $\lambda$ is a closed $n$-form on $\mathcal{Z}$ (or a section $\lambda$ of $\Lambda^nT^* \mathcal{Z}\longrightarrow \mathcal{Z}$ s.t. $\lambda^*\omega = 0$) which is a solution of $\lambda^*\mathcal{H} =0$.

Then if for instance $\mathcal{Z} = \mathcal{X}\times \mathcal{Y}$, and if we denote by $\pi$ the projection from $\mathcal{X}\times \mathcal{Y}$ to $\mathcal{X}$, $\lambda:= \pi^*dS$ provides us with a \emph{null Lagrangian functional} $\int_\mathcal{X}\lambda$ on $\mathcal{X}$ (i.e. a Lagrangian density which satisfies the Euler--Lagrange equation for any map). In constrast with non relativistic quantum Mechanics, the usefulness of this equations in the quantization of fields is not clear for the moment. One of the interests of the Hamilton--Jacobi equation is that it allows in principle to prove under some circumstances that some solutions of the Euler--Lagrange system of equations are glogal minimizers, by following a classical strategy designed by K. Weierstrass and D. Hilbert (see \cite{Weyl,Dedecker,Rund}). This strategy is the exact analogue in the general theory of calculus of variations of the theory of \emph{calibrations} used in minimal surfaces. 

Note that one could impose extra conditions such as requiring that $\lambda = ds^1\wedge \cdots \wedge ds^n$, where $s^1,\cdots, s^n$ are functions on $\mathcal{Z}$ plus the fact that the graph of $\lambda$ is foliated by solutions to the Hamilton equations (this provides then a generalization of the picture built by Hamilton in order to conciliate the Fermat principle with the Huygens principle): this was achieved by Carath{\'e}odory \cite{caratheodory} in his theory (see \S \ref{paragraph1.8}).

\subsection{Some historical remarks}\label{paragraph1.8}
The generalization of the Hamilton equations to variational problems with several variables developped first along two directions. One of these is the question of deciding whether a given solution to a variational problem is a minimum of the action functional. This question was answered locally for 1-dimensional variational problems by C.G.J. Jacobi (by following a remark of Legendre) in 1837 \cite{jacobi-legendre} by founding a method to check that the second variation is nonnegative which is based on solutions to the so-called Jacobi equation. Note that this method was extended to several variables by A. Clebsch \cite{clebsch} in 1859. Later on a global, nonlinear version of these ideas was developped by K. Weierstrass and D. Hilbert to prove the minimality of some solutions. This theory is connected with another famous work of Jacobi of the same year (1837), who obtained the Hamilton--Jacobi equation \cite{jacobi-hamilton} by generalizing the work of Hamilton relating the Fermat principle to the Huygens principle. In 1890 V. Volterra wrote two papers \cite{volterra1,volterra2} where, to my knowledge for the first time\footnote{This was followed by a work by L. Koenigsberger \cite{koenigsberger} in 1901, quoted by T. De Donder in \cite{DeDonder}, which unfortunately I have difficulties to understand.}, two different generalizations of the Hamilton system of equations to variational problems with several variables were proposed. In \cite{volterra2} Volterra extended the Weierstrass--Hilbert theory to variational problems with several variables. This theory was further developped by G. Prange in 1915 \cite{prange} and by C. Carath{\'e}odory in his book in 1929 \cite{caratheodory} and is called today the \emph{Carath{\'e}odory} theory. In 1934 H. Weyl \cite{Weyl}, inspired by Carath{\'e}odory's theory, proposed a variant of it which is based on the same theory as the one proposed by Volterra in \cite{volterra1} and that we described in \S \ref{paragraph1.1}. Today this theory is called the \emph{De Donder--Weyl} one by many Authors\footnote{including, in previous papers, the Author of this note, who ignored until recently the work of Volterra}. Its geometrical framework is the affine multisymplectic manifold $\Lambda^n_2T^*\mathcal{Z}$.

A second direction was the notion of \emph{invariant integrals} due to H. Poincar{\'e} \cite{poincare} and further developped by E. Cartan \cite{cartan} in 1922: here one emphasizes the relationship of Hamilton equations with the search of invariants which may be functions or differential forms. This point of view is strongly related with the covariant phase space theory (see \S \ref{historical2} below). This theory was developped in full generality by T. De Donder \cite{DeDonder} in 1935 and his main contribution was to deduce the extension of the affine (`De Donder--Weyl') theory to Lagrangian densities depending on an arbitrary number of derivatives, i.e. to the theory expounded in \S \ref{paragraph1.2}. Hence although Weyl's and De Donder's contributions are almost simultaneous they are independant in their inspiration: Weyl starting point was the so-called Carath{\'e}odory theory, motivated by the search for generalizations to several variables of the Hamilton--Jacobi equation, whereas De Donder's starting point was the theory of integral invariants.

The fact that a continuum of different theories may exists for a given variational problem was first understood by T. Lepage \cite{Lepage} in 1936 and completely described by P. Dedecker in 1953 \cite{Dedecker}. Today we can picture these various theories as submanifolds of the universal multisymplectic manifold $\Lambda^nT^*\mathcal{Z}$ introduced by J. Kijowski \cite{Kijowski2} in 1974.

Recently the so-called De Donder--Weyl theory (but that we should call the `first Volterra theory') has beed studied by many Authors starting with the important work by the Polish school around 1970, i.e. by W. Tulczjew, J. Kijowski, W. Szczyrba and later on in many papers which are refered to in e.g. \cite{helein,forgergomes}. However the Lepage--Dedecker theory has received much less attention (to my knowledge it was only considered by J. Kijowski \cite{Kijowski2}, F. H{\'e}lein, J. Kouneiher \cite{hk1,hk2,helein} and M. Forger, S. V. Romero \cite{forgerromero}), probably because of its complexity. The latter theory leads however to interesting phenomena, particularly for gauge theories \cite{hk1,helein}, since first class Dirac constraints simply disappear there.

The modern formulation using the multisymplectic $(n+1)$-form as the key of the structure of the theory seems to start with the papers of J. Kijowski \cite{Kijowski1}, H. Goldschmidt and P. Sternberg \cite{goldstern} in 1973 and the introduction of observable $(n-1)$-forms apparentely goes back to the work of K. Gaw{\'e}dski \cite{Gawedski} in 1972.

\subsection{An example}\label{paragraph-example}
Let $\mathcal{X}$ be the $n$-dimensional Minkowski space-time with coordinates $x  = (x^0,x^1,\cdots ,x^{n-1})$ and consider the linear Klein--Gordon equation on $\mathcal{X}$:
\begin{equation}\label{kg}
\square \varphi + m^2\varphi := {\partial^2\varphi \over \partial t^2} -
 \Delta \varphi + m^2\varphi = 0,
\end{equation}
where $t=x^0$ and $\Delta:= \sum_{i=1}^{n-1}\frac{\partial^2}{(\partial x^i)^2}$. We use the notations $\vec{x}:=
(x^1,\cdots ,x^{n-1})\in \mathbb{R}^{n-1}$ and $x= (x^0,\vec{x})\in \mathbb{R}^n$ and we define the Euclidean scalar product $\vec{x}\cdot \vec{y}:= x^1y^1 + \cdots + x^{n-1}y^{n-1}$ on $\mathbb{R}^{n-1}$ and the Minkowski product 
\[
x\cdot y = \eta_{\mu\nu}x^\mu y^\nu = x^0y^0 - x^1y^1 - \cdots -
x^{n-1}y^{n-1} = x^0y^0 - \vec{x}\cdot \vec{y}, 
\]
on $\mathcal{X}$. The multisymplectic formulation of (\ref{kg}) takes place in $\mathcal{M}:= \Lambda^nT^*(\mathcal{X}\times \R)$, equipped with the multisymplectic form
\[
\omega:= de\wedge \beta + dp^\mu\wedge d\varphi\wedge \beta_\mu.
\]
Note that $\omega = d\theta_{\lambda}$, where
\[
\theta_{\lambda}:= e \beta + \lambda p^\mu d\varphi\wedge \beta_\mu
- (1-\lambda)\varphi dp^\mu\wedge \beta_\mu,
\]
where $\lambda\in \R$ is a parameter to fixed later. The Hamiltonian function on $\mathcal{M}$ corresponding to solutions of (\ref{kg}) is
\[
\mathcal{H}(x,\varphi,e,p):= e + {1\over 2}\eta_{\mu\nu}p^\mu p^\nu +
{1\over 2} m^2\varphi^2.
\]
To a solution $\varphi$ of (\ref{kg})we associate a Hamiltonian $n$-curve $\Gamma = \{(x,\varphi(x),e(x),p(x))\ |\ x\in \mathcal{X}\}$ which satisfies
\begin{equation}\label{pe}
\left\{\begin{array}{ccl}
p^\mu(x) & = & \displaystyle \eta^{\mu\nu}{\partial \varphi\over \partial
  x^\nu}(x)\\
e(x) & = &\displaystyle  - {1\over 2}\eta^{\mu\nu}{\partial
  \varphi\over \partial x^\mu}(x){\partial \varphi\over \partial
  x^\nu}(x) - {1\over 2}m^2\varphi(x)^2. 
\end{array}\right.
\end{equation}
We define $\mathcal{E}$ to be the set of Hamiltonian $n$-curves s.t. for all time $t$, $\vec{x}\longmapsto \varphi(t,\vec{x})$ is rapidly decreasing at infinity.
% we define the subset $\mathcal{L}(\mathcal{E})$ of "linear" functionals on $\mathcal{E}$.

We denote by $\gP^{n-1}_\mathcal{H}\mathcal{M}$ the set of dynamical observable $(n-1)$-forms $F$ and ${\gsp}_\mathcal{H}\mathcal{M}:= \{\xi\ |\
L_\xi \omega = 0, d\mathcal{H}(\xi) = 0\}$. Note that $(n-1)$-forms $F$ in $\gP^{n-1}_\mathcal{H}\mathcal{M}$ are found by looking at vector fields $\xi$ in ${\gsp}_\mathcal{H}\mathcal{M}$ and by solving $\xi\iN \omega + dF = 0$. They are of the form
\[
F = \zeta\iN \theta + F_\Phi,
\]
where $\zeta$ is a vector field on the Minkowski space-time $\mathcal{X}$
which is a generator of the action of the Poincar{\'e} group and 
\[
F_\Phi:= \left( p^\mu\Phi(x) -\varphi\eta^{\mu\nu}{\partial
    \Phi\over\partial x^\nu}(x) 
\right) \beta_\mu,
\]
where $\Phi$ is a solution of (\ref{kg}). Note that moreover
\[
\xi_\Phi:=\xi_{F_\Phi} = \Phi(x){\partial \over \partial \varphi} + \eta^{\mu\nu}{\partial \Phi\over\partial x^\nu}(x)
{\partial \over \partial p^\mu} - \left( m^2\varphi\Phi(x) + p^\mu
{\partial \Phi\over\partial x^\mu}(x)\right) {\partial \over \partial e}.
\]
In the following we shall denote by
\[
P_\mu^{(\lambda)}:= {\partial \over \partial x^\mu}\iN \theta_{\lambda}
\]
and we observe that since $L_{\partial \over \partial x^\mu}\theta_\lambda = 0$, we have $dP_\mu^{(\lambda)} + {\partial \over \partial x^\mu}\iN\omega =0$. Hence $\xi_{P_\mu^{(\lambda)}} = \frac{\partial}{\partial x^\mu}$.

The brackets of two dynamical observable forms $F,G\in \gP^{n-1}_\mathcal{H}\mathcal{M}$ are given as follows:
for any pair $\Phi$, $\Psi$ of solutions of (\ref{kg}), 
\begin{equation}\label{bracketphipsi}
\{F_\Phi,F_\Psi\} = \eta^{\mu\nu}\left(
{\partial \Phi\over \partial x^\nu}(x)\Psi(x) -
\Phi(x){\partial \Psi\over \partial x^\nu}(x)\right)\beta_\mu.
\end{equation}
We observe that $d\{F_\Phi,F_\Psi\} = 0$. Hence
$\left(\gP^{n-1}_\mathcal{H}\mathcal{M}, \{\cdot ,\cdot\}\right)$ can be understood as a kind of central extension of the Lie algebra $\left( {\gsp}_\mathcal{H}\mathcal{M}, [\cdot ,\cdot ]\right)$ and the Lie sub algebra spanned by forms $F_\Phi$ as an infinite dimensional analogue of the Heisenberg algebra with central charges given by (\ref{bracketphipsi}). Lastly 
\begin{equation}\label{pf}
\left\{P_\mu^{(\lambda)},F_\Phi\right\} = L_{\partial \over \partial
  x^\mu}F_\Phi - d\left( {\partial \over \partial x^\mu}\iN
  F_\Phi\right) = F_{\partial \Phi\over \partial x^\mu} -
d\left( {\partial \over \partial x^\mu}\iN F_\Phi\right)
\end{equation}
and $\left\{P_\mu^{(\lambda)},P_\nu^{(\lambda)}\right\} = 0$.

For the purpose of quantization we look at functionals of the form $\mathcal{F} = \int_\Sigma F_\Phi$ which are simultaneously eigenvectors of the linear operators
\[
\mathcal{F} \longmapsto \left\{\int_\Sigma P_\mu^{(\lambda)},\mathcal{F}\right\},
\]
for $\mu = 0,\cdots ,n-1$. We find by using relation (\ref{pf}) that the eigenvector equation reduces to ${\partial \Phi\over \partial x^\lambda} =
c\Phi$. This implies (by using the eigenvalue equation for $\mu =
0,\cdots ,n-1$) that $\Phi(x) = \alpha e^{ik\cdot x}$. But because
$\Phi$ should also be a solution of (\ref{kg}) we must then have
\begin{equation}\label{shell}
\eta_{\mu\nu}k^\mu k^\nu = (k^0)^2 - |\vec{k}|^2 = m^2.
\end{equation}
(We remark that the maps $\{\int_\Sigma P_\lambda^{(\lambda)},\cdot\}$ play
the role of the generators of a Cartan subalgebra.)  Let us denote by
$\mathcal{C}$ the \emph{mass shell}, i.e. the set of all $k = (k^0,\vec{k})\in \R^4$ which are
solutions of (\ref{shell}). This set actually splits into two connected components
according to the sign 
of $k^0$: we let $\mathcal{C}^+:= \{k\in \mathcal{C}\ |\ k^0>0\}$. For any
$k\in \mathcal{C}^+$ we define  
\[
\begin{array}{ccccl}
\alpha_k & := & \displaystyle F_{ie^{ik\cdot x}/\sqrt{2\pi}^3} & = &
\displaystyle {i\over \sqrt{2\pi}^3} e^{ik\cdot x}\left( p^\mu -i
  \varphi k^\mu\right)\beta_\mu \\
\alpha^*_k & := & \displaystyle F_{-ie^{-ik\cdot x}/\sqrt{2\pi}^3} & =
& \displaystyle {-i\over \sqrt{2\pi}^3} e^{-ik\cdot x}\left( p^\mu +i
  \varphi k^\mu\right)\beta_\mu. 
\end{array}
\]
The vector fields associated to these observable forms are:
\[
\xi_k:= \xi_{\alpha_k} = {e^{ik\cdot x}\over \sqrt{2\pi}^3}\left[
  i{\partial \over \partial \varphi} - k^\mu {\partial \over \partial
  p^\mu} + \left( \eta_{\mu\nu}p^\mu k^\nu - im^2\varphi\right)
  {\partial \over \partial e} \right],
\]
\[
\xi^*_k:= \xi_{\alpha^*_k} = {e^{-ik\cdot x}\over \sqrt{2\pi}^3}\left[
  -i{\partial \over \partial \varphi} - k^\mu {\partial \over \partial
  p^\mu} + \left( \eta_{\mu\nu}p^\mu k^\nu + im^2\varphi\right)
  {\partial \over \partial e} \right].
\]
We then define the observable functionals
\[
a_k :=  \int_\Sigma \alpha_k \quad \hbox{ and }\quad  a^*_k :=
\int_\Sigma \alpha^*_k. 
\]
As the notations suggest these functionals are the classical analogues of respectively
the annihilation and the creation operators. The advantage
however is that our functionals $a_k$ and $a^*_k$ are independant of
the coordinate system. We can choose $\Sigma$ to be the hyperplane
$x^0=t=0$ and, for any function $f$, denote by $f|_0$ the restriction of $f$ to $\Sigma$. Then, for any $\Gamma\in \mathcal{E}$ we have 
\[
%\begin{array}
%{ccccl}& \displaystyle 
a_k(\Gamma)  = {i\over
  \sqrt{2\pi}^3}\int_{\R^3}\left(
{\partial \varphi \over\partial t}|_0(\vec{x}) - ik^0\varphi|_0(\vec{x})\right)
e^{-i\vec{k}\cdot\vec{x}} d\vec{x} 
 =  i\widehat{{\partial \varphi\over\partial
    t}|_0}(\vec{k}) + k^0\widehat{\varphi|_0}(\vec{k}), 
%\end{array}
\]
where, for all function $\psi$ on $\R^3$,
\[
\widehat{\psi}(\vec{k}):= {1\over
  \sqrt{2\pi}^3}\int_{\R^3}\psi(\vec{x})e^{-i\vec{k}\cdot\vec{x}}
d\vec{x}. 
\]
Similarly we have:
\[
a^*_k(\Gamma) = {-i\over
  \sqrt{2\pi}^3}\int_{\R^3}\left({\partial \varphi \over\partial
    t}|_0(\vec{x}) + ik^0\varphi|_0(\vec{x})\right)
e^{i\vec{k}\cdot\vec{x}} d\vec{x} 
=  -i\widehat{{\partial \varphi\over\partial
  t}|_0}(-\vec{k}) + k^0\widehat{\varphi|_0}(-\vec{k}). 
\]
Hence we deduce that
\[
\widehat{\varphi|_0}(\vec{k}) = {1\over 2k^0}\left(
  a_k(\Gamma) + a^*_{\overline{k}}(\Gamma)\right) 
\quad \hbox{and} \quad \widehat{{\partial \varphi \over\partial t}|_0}(\vec{k}) = {1\over 2i}\left( a_k(\Gamma) -
  a^*_{\overline{k}}(\Gamma)\right), 
\]
where $\overline{k} = \overline{(k^0,\vec{k})}:= (k^0,-\vec{k})$. Thus denoting $d\mu(k) = {1\over 2k^0}dk^1dk^2dk^3 = {1\over 2k^0}d\vec{k}$, we have
\[
\begin{array}{ccl}
\varphi(0,\vec{x}) & = & \displaystyle {1\over \sqrt{2\pi}^3}\int_{\R^3}{1\over 2k^0}d\vec{k}\ e^{i\vec{k}\cdot \vec{x}}\left(a_k(\Gamma) + a^*_{\overline{k}}(\Gamma)\right)\\
& = & \displaystyle {1\over \sqrt{2\pi}^3}\int_{\mathcal{C}^+}d\mu(k) \left(a_k(\Gamma) e^{-ik\cdot x} + a^*_k(\Gamma)e^{ik\cdot x}\right)
  \end{array}
\]
and
\[
\begin{array}{ccl}
{\partial \varphi \over\partial t}(0,\vec{x}) & = & \displaystyle {-i\over  \sqrt{2\pi}^3}\int_{\R^3}{1\over 2}d\vec{k}\ e^{i\vec{k}\cdot \vec{x}}\left(a_k(\Gamma) - 
  a^*_{\overline{k}}(\Gamma)\right) \\
& = & \displaystyle {-i\over
  \sqrt{2\pi}^3}\int_{\mathcal{C}^+}d\mu(k)k^0 \left(a_k(\Gamma)
  e^{-ik\cdot x} - a^*_k(\Gamma)e^{ik\cdot x}\right). 
  \end{array}
\]
Recall that these integrals can be interpreted as integrals over
$\mathcal{C}^+$ through the parametrization $\R^3\ni \vec{k}\longmapsto
(k^0,\vec{k})\in \mathcal{C}^+$ and that $d\mu$ is a
measure on $\mathcal{C}^+$ invariant by the action of the Lorentz
group. Note also that in order to agree with some textbooks one should
add an extra factor $\sqrt{k^0}$ inside the integrals. By using the
relation (\ref{bracketphipsi}) we obtain, $\forall k,\ell \in {\cal
  C}^+$, 
\[
\{\alpha_k, \alpha_{\ell}\} = {-ie^{i(k+\ell)\cdot x}\over
  (2\pi)^3}(k^\mu - \ell^\mu)\beta_\mu, 
\]
\[
\{\alpha^*_k, \alpha^*_{\ell}\} = {ie^{-i(k+\ell)\cdot x}\over
  (2\pi)^3}(k^\mu - \ell^\mu)\beta_\mu, 
\]
\[
\{\alpha_k, \alpha^*_{\ell}\} = {ie^{i(k-\ell)\cdot x}\over
  (2\pi)^3}(k^\mu + \ell^\mu)\beta_\mu. 
\]
These brackets cannot be integrated over the slice $\Sigma:=\{x^0 =
0\}$ in the measure theoretical sense\footnote{in contrast with the
  integrals $\int_{\Sigma\cap \Gamma} \alpha_k$ and $\int_{\Sigma\cap
    \Gamma} \alpha^*_k$ which exists if the restrictions to $\Sigma$
  of $\varphi$ and of its time derivative are Lebesgue integrable},
but one can make sense of their integrals as distributions over the
variables $\vec{k}\pm \vec{\ell}\in \R^3$: 
\[
\{a_k,a_{\ell}\} = \{a^*_k,a^*_{\ell}\} = 0,\quad \forall k,\ell\in \mathcal{C}^+,
\]
and
\[
\{a_k,a^*_{\ell}\} = i2k^0 \delta(\vec{\ell}-\vec{k}).
\]
A way to regularize these operators and their brackets is, by using
functions $f,g\in L^2(\mathcal{C}^+)$, to define 
\[
a_f:= \int_{\mathcal{C}^+} d\mu(k)f(k) a_k,
\quad \hbox{and}\quad
a^*_g:= \int_{\mathcal{C}^+}  d\mu(k)g(k) a^*_k.
\]
Then
\[
\{a_f,a^*_g\} = i\int_{\mathcal{C}^+} d\mu(k)f(k)g(k).
\]

\section{The covariant phase space}

\subsection{A short historical review}\label{historical2}
The simplest version of the covariant phase space is the set $\mathcal{E}$ of solutions of a Hamiltonian time evolution problem. In this case the Cauchy problem consists in choosing some point $\textsc{m}_0$ in the ordinary phase space (classically positions and momenta) and some time $t_0$ and in looking for solutions of the Hamiltonian vector flow which coincide with $\textsc{m}_0$ at time $t_0$. This problem has an unique solution in all regular cases and this means that $\mathcal{E}$ is in one to one correspondence with the set of initial data $\{\textsc{m}_0\}$. In other words to each time $t_0$ it corresponds a natural `Cauchy coordinates system' on $\mathcal{E}$, which is just the set of initial conditions $\{\textsc{m}_0\}$. The key property is that the Hamiltonian flow preserves the symplectic structure: this means all the symplectic structures induced by these `Cauchy coordinates systems' on $\mathcal{E}$ coincides. Hence this defines a canonical symplectic structure on $\mathcal{E}$. The substitution of the ordinary phase space by the space of solutions is a classical analogue of the transition from the \emph{Schr{\"o}dinger picture} to the \emph{Heisenberg picture} in quantum Mechanics: in the Schr{\"o}dinger picture the dynamics of a particle is described by the evolution of some time dependant `state' which is represented by a complex line in some complex Hilbert space (the quantum analogue of a point in the ordinary phase space), whereas in the Heisenberg picture the state (still a complex line in a complex Hilbert space) does not evolve with time so that it may be interpreted as a quantum analogue of a solution of the dynamical equations, i.e. of a point in $\mathcal{E}$ (actually more precisely on a Lagrangian submanifold in the phase sapce, according to A. Weinstein).

In Mechanics this concept is relatively old: the idea of contempling the space of solutions of a mechanical problem itself has probably his roots in the method of the \emph{variation de la constante} of J.L. Lagrange and the notion of `Lagrange bracket' is very close to the symplectic structure on the phase space. The observation that this space carries an intrinsic symplectic structure was clearly formulated by H. Poincar{\'e} \cite{poincare} in his theory of \emph{invariant integrals} (\emph{invariants int{\'e}graux}) and later further developped by E. Cartan \cite{cartan} and fully recognized by J.M. Souriau \cite{souriau}.  T. De Donder \cite{oldDD} extended the notion of integral invariant to variational problems with several variables, being hence very close from the notion of covariant phase space in this context, although it is not clear that he realized it. Actually it seems difficult to decide when the concept of covariant phase space in fields theory merged out. My own guess is that such an idea could also have been inspired by quantum fields theory, since it may be thought as the classical analogue of the Heisenberg picture in quantum fields theory. First known works in this direction are the R.E. Peierls bracket in 1952 \cite{peierls}, followed by the paper of I. Segal \cite{segal} in 1960. Peierls defined a bracket on the set of solutions to a relativic hyperbolic wave equation which can be understood now as the restriction of the Poisson bracket associated to the covariant phase structure on a certain class of functionals on the phase space. Segal proved that the set of solutions of a non linear field relativistic wave equation precisely carries a symplectic structure and proposes to use this symplectic structure to quantize fields (and his paper is among the ones at the origin of the geometric quantization method). This idea was later developped in a more and more general framework by P. L. Garc{\'\i}a \cite{garcia} in 1968, Garc{\'\i}a and A. P{\'e}rez-Rend{\'o}n \cite{garcia-perez-rendon} in 1971, H. Goldschmidt, S. Sternberg \cite{goldstern} in 1973. To my opinion the more achieved presentation is the one by J. Kijowski and W. Szczyrba \cite{KijowskiSzczyrba} in 1976, which gives the first elementary but general presentation of this structure, by using the multisymplectic formalism.

A more recent apparition of this idea can be found in the papers by C. Crnkovic and E. Witten \cite{CrnkovicWitten} and by G. Zuckerman \cite{zuckerman}, where the Authors apparentely ignored the previous references and have rediscovered this principle, being guided by the concept of the variational bicomplex by F. Takens \cite{takens} and from the work of A.M. Vinogradov \cite{vinogradov}. This was followed by several developments in the physical (e.g. \cite{DolanHaugh}) and the mathematical literature, where this principle is often refered to as the Witten covariant phase space. A general presentation in the framework of the \emph{secondary calculus} of Vinogradov was done by E. Reyes \cite{reyes} and L. Vitagliano in \cite{Vitagliano} and in relation to multisymplectic geometry (as in the present paper) by M. Forger and S.V. Romero in \cite{forgerromero}.

\subsection{The basic principle}
We expound here briefly the principle of the covariant phase space using the multsymplectic formalism. Our presentation will be heuristic and we refer to \cite{KijowskiSzczyrba,heleinprep} for details. We assume that we are given a premultisymplectic manifold $(\mathcal{M}, \omega,\beta)$ (see \S \ref{paragraph1.4}) and, as in \S\ref{paragraph1.5}, that $\omega$ is exact, i.e. $\omega = d\theta$, for some $n$-form $\theta$. We note $\mathcal{E}$ the set of Hamiltonian $n$-curves in $(\mathcal{M}, \omega,\beta)$, i.e. the set of oriented $n$-dimensional submanifolds $\Gamma\subset \mathcal{M}$ which satisfy (\ref{Hamilton3}). Given some $\Gamma\in \mathcal{E}$, the tangent space\footnote{Note that since $\mathcal{E}$ may not be a manifold in general, the usual definition of a tangent space should be replaced by a suitable notion, see \cite{KijowskiSzczyrba,heleinprep}} to $\mathcal{E}$ at $\Gamma$ represents the set of infinitesimal deformations $\delta \Gamma$ of $\Gamma$ which preserves the equation (\ref{Hamilton3}). Such a deformation $\delta \Gamma$ can be represented by a vector field $\xi$ tangent to $\mathcal{M}$ defined along $\Gamma$, i.e. a section over $\Gamma$ of $j_\Gamma^*T\mathcal{M}$, which is the pull-back image of the tangent bundle $T\mathcal{M}$ by the embedding map $j_\Gamma:\Gamma\longrightarrow \mathcal{M}$. Given $\delta\Gamma$, the vector field $\xi$ is of course not unique, since for any tangent vector field $\zeta$ on $\Gamma$ (i.e. a section of the subbundle $T\Gamma \subset j_\Gamma^*T\mathcal{M}$), $\xi+\zeta$ represents also $\delta \Gamma$. If so we write:
\[
\delta \Gamma = \int_\Gamma \xi = \int_\Gamma \xi + \zeta.
\]
Moreover the condition on $\delta \Gamma$ of being tangent to $\mathcal{E}$ forces $\xi$ to be a solution of the \emph{Jacobi equation}:
\begin{equation}\label{Jacobi}
\forall v\in \mathcal{C}^\infty(\mathcal{M},T_\textsc{m}\mathcal{M}),\quad
\left(v\iN L_\xi\omega\right)|_\Gamma = 0.
\end{equation}
Note that, although $\xi$ is not a vector field defined on $\mathcal{M}$ (neither on a neighbourhood of $\Gamma$ in $\mathcal{M}$) but only on $\Gamma$, one can make sense of $L_\xi\omega|_\Gamma$ because $\Gamma$ is a solution of (\ref{Hamilton3}).

Then for any slice $\Sigma$ (see \S \ref{paragraph1.6}), any $\Gamma\in \mathcal{E}$ and $\delta \Gamma\in T_\Gamma\mathcal{E}$, we define
\[
\Theta^\Sigma_\Gamma(\delta\Gamma):= \int_{\Sigma\cap\Gamma}\xi\iN \theta,
\]
where $\xi$ is a section of $j_\Gamma^*T\mathcal{M}$ over $\Gamma$ s.t. $\delta\Gamma = \int\xi$ and $\xi\iN\theta$ is the interior product of $\theta$ by $\xi$. This hence define a 1-form $\Theta^\Sigma$ on $\mathcal{E}$

\subsubsection{The dependance of $\Theta^\Sigma$ on $\Sigma$}

This is the first natural question. For that purpose let us consider a smooth 1-parameter family of slices $(\Sigma_t)_t$ and compute the derivative:
\[
\begin{array}{ccl}
\displaystyle \frac{d}{dt}\left( \Theta^{\Sigma_t}_\Gamma(\delta\Gamma)\right)&  = & \displaystyle \frac{d}{dt}\left(\int_{\Sigma_t\cap\Gamma}\xi\iN \theta\right) = \int_{\Sigma_t\cap\Gamma}L_{\frac{\partial}{\partial t}} (\xi\iN \theta)\\
& = & \displaystyle \int_{\Sigma_t\cap\Gamma} \frac{\partial}{\partial t}\iN d(\xi\iN \theta) + d\left(\xi\wedge \frac{\partial}{\partial t}\iN\theta\right).
\end{array}
\]
But $d(\xi\iN \theta) = L_\xi\theta - \xi\iN d\theta$ and thus
\[
\frac{\partial}{\partial t}\iN d(\xi\iN \theta) = \frac{\partial}{\partial t}\iN \left(L_\xi\theta\right) - \frac{\partial}{\partial t}\iN \xi\iN\omega.
\]
However we can assume w.l.g. (see \cite{heleinprep}) that the vector fields $\frac{\partial}{\partial t}$ and $\xi$ admit extensions s.t. $\left[\xi,\frac{\partial}{\partial t}\right] = 0$. Then the preceding relation gives us
\[
\frac{\partial}{\partial t}\iN d(\xi\iN \theta) = L_\xi\left(\frac{\partial}{\partial t}\iN \theta\right) - \xi\wedge \frac{\partial}{\partial t}\iN\omega.
\]
Hence
\begin{equation}\label{calcul1}
\frac{d}{dt}\left( \Theta^{\Sigma_t}_\Gamma(\delta\Gamma)\right) =
\int_{\Sigma_t\cap\Gamma} L_\xi\left(\frac{\partial}{\partial t}\iN \theta\right) - \int_{\Sigma_t\cap\Gamma} \xi\wedge \frac{\partial}{\partial t}\iN\omega + \int_{\Sigma_t\cap\Gamma} d\left(\xi\wedge \frac{\partial}{\partial t}\iN\theta\right).
\end{equation}
First let us consider a smooth curve $s\longmapsto \Gamma_s\in \mathcal{E}$ s.t. $\Gamma_0 = \Gamma$ and $\frac{d\Gamma_s}{dt} = \delta \Gamma$. Then the first term in the r.h.s. of (\ref{calcul1}) is equal to
\[
\int_{\Sigma_t\cap\Gamma} L_\xi\left(\frac{\partial}{\partial t}\iN \theta\right) = \frac{d}{ds}\left. \left(\int_{\Sigma_t\cap\Gamma_s}\frac{\partial}{\partial t}\iN \theta\right)\right|_{s=0} =
\delta S^{\frac{d\Sigma_t}{dt}}_\Gamma(\delta\Gamma),
\]
where we have posed:
\[
S^{\frac{d\Sigma_t}{dt}}(\Gamma):= \int_{\Sigma_t\cap\Gamma}\frac{\partial}{\partial t}\iN \theta.
\]
Second we can assume w.l.g. (see \cite{heleinprep}) that we can choose $\frac{\partial}{\partial t}$ in such a way that it is tangent to $\Gamma$. Let $(X_2,\cdots , X_n)$ be a system of tangent vectors on $\Gamma$ s.t. $\forall t$, $\forall \textsc{m}\in \Sigma_t\cap\Gamma$, $(X_2(\textsc{m}),\cdots , X_n(\textsc{m}))$ is a basis of $T_\textsc{m}(\Sigma_t\cap\Gamma)$ and $(\frac{\partial}{\partial t}(\textsc{m}),X_2(\textsc{m}),\cdots , X_n(\textsc{m}))$ is a basis of $T_\textsc{m}\Gamma$. Then if $\psi$ is a $n$-volume form on $\Gamma$ s.t. $\psi(\frac{\partial}{\partial t},X_2,\cdots , X_n) = 1$, the second term in the r.h.s. of (\ref{calcul1}) reads
\[
- \int_{\Sigma_t\cap\Gamma} \xi\wedge \frac{\partial}{\partial t}\iN\omega = - \int_{\Sigma_t\cap\Gamma} \omega\left(\xi,\frac{\partial}{\partial t},X_2,\cdots , X_n\right)\psi
\]
and vanishes because of the Hamilton equations (\ref{Hamilton3}). Lastly we assume that \emph{the restriction of $\xi$ to $\Sigma_t\cap\Gamma$ has compact support or is rapidly decreasing}: this occurs for instance if the Hamilton system encodes hyperbolic wave equations, if $\Sigma$ is a level hypersurface of some time coordinate and if we impose that the Hamiltonian $n$-curves in $\mathcal{E}$ have a prescribed behaviour at infinity in space for all time. Then the last term in the r.h.s. of (\ref{calcul1}) vanishes. Then Relation (\ref{calcul1}) can be rewritten 
\[
\frac{d}{dt}\left( \Theta^{\Sigma_t}_\Gamma(\delta\Gamma)\right) =
\delta S^{\frac{d\Sigma_t}{dt}}_\Gamma(\delta\Gamma), \quad \forall \delta\Gamma\in T_\Gamma\mathcal{E}
\]
or
\begin{equation}\label{dthetasurdt}
\frac{d}{dt}\left( \Theta^{\Sigma_t}\right) = \delta S^{\frac{d\Sigma_t}{dt}}.
\end{equation}
We can also define the functional
\[
S^{\Sigma_2}_{\Sigma_1}(\Gamma):= \int_{\Gamma\cap \{t_1\leq t\leq t_2\}}\theta,
\]
which represents the `action' between the slices $\Sigma_1:= \{t=t_1\}$ an $\Sigma_2:= \{t=t_2\}$. Then $S^{\Sigma_2}_{\Sigma_1}(\Gamma) = \int_{t_1}^{t_2}S^{\frac{d\Sigma_t}{dt}}(\Gamma)dt$ and thus we deduce by integrating (\ref{dthetasurdt}) over $[t_1,t_2]$ that
\begin{equation}\label{differencedeTheta}
\Theta^{\Sigma_2} - \Theta^{\Sigma_1} = \delta S^{\Sigma_2}_{\Sigma_1}.
\end{equation}

\subsubsection{The symplectic form}

In view of the preceding we are led to the conclusion that, although the 1-form $\Theta^\Sigma$ depends on $\Sigma$, its differential $\delta\Theta^\Sigma$ does not depend on $\Sigma$ since (\ref{differencedeTheta}) tells us that $\Theta^{\Sigma_2} - \Theta^{\Sigma_1}$ is an exact form. Of course one should be careful in using the identity $\delta \circ \delta =0$ since $\mathcal{E}$ is not a smooth manifold (see \cite{heleinprep} for a rigorous proof that $\delta \Theta^\Sigma$ does not depend on $\Sigma$). All that motivates the definition of the following 2-form on $\mathcal{E}$:
\[
\Omega := \delta\Theta^\Sigma.
\]
We will prove that $\Omega$ has the following expression: $\forall \delta_1\Gamma,\delta_2\Gamma \in T_\Gamma\mathcal{E}$,
\begin{equation}\label{deltaTheta=Omega}
\Omega_\Gamma(\delta_1\Gamma,\delta_2\Gamma) = \int_{\Sigma\cap\Gamma} \xi_1\wedge \xi_2\iN\omega,
\end{equation}
where $\xi_1,\xi_2$ are sections over $\Gamma$ of $j_\Gamma^*T\mathcal{M}$ s.t. $\delta_1\Gamma = \int_\Gamma \xi_1$ and $\delta_2\Gamma = \int_\Gamma \xi_2$. To prove (\ref{deltaTheta=Omega}) we need to compute $\delta\Theta^\Sigma_\Gamma(\delta_1\Gamma,\delta_2\Gamma)$. For that purpose we first assume that we can extend the two tangent vectors $\delta_1\Gamma$ and $\delta_2\Gamma$ to commuting vector fields on $\mathcal{E}$ around $\Gamma$ (actually we can assume that $[\xi_1,\xi_2]=0$). Then
\[
\begin{array}{ccl}
\delta \Theta^\Sigma_\Gamma(\delta_1\Gamma,\delta_2\Gamma) & = & \delta_1\Gamma\cdot \Theta^\Sigma_\Gamma(\delta_2\Gamma) - \delta_2\Gamma\cdot \Theta^\Sigma_\Gamma(\delta_1\Gamma) - \Theta^\Sigma_\Gamma([\delta_1\Gamma,\delta_2\Gamma])\\
& = & \displaystyle \delta_1\Gamma\cdot \left(\int_{\Sigma\cap\Gamma}\xi_2\iN \theta\right)  - \delta_2\Gamma\cdot \left(\int_{\Sigma\cap\Gamma}\xi_1\iN \theta\right).
\end{array}
\]
Thus
\[
\delta \Theta^\Sigma_\Gamma(\delta_1\Gamma,\delta_2\Gamma) = \int_{\Sigma\cap\Gamma} L_{\xi_1}\left(\xi_2\iN \theta\right) - \int_{\Sigma\cap\Gamma} L_{\xi_2}\left(\xi_1\iN \theta\right).
\]
We use then the following identity (see \cite{heleinprep}): for any pair of vector fields $X_1$ and $X_2$ and for any $p$-form $\beta$,
\[
L_{X_1}(X_2\iN \beta) - L_{X_2}(X_1\iN \beta) = X_1\wedge X_2\iN d\beta + [X_1,X_2]\iN\beta 
+ d(X_1\wedge X_2\iN \beta).
\]
Setting $X_1 = \xi_1$, $X_2=\xi_2$ and $\beta = \theta$, we obtain using $[\xi_1,\xi_2]= 0$ and $d\theta =\omega$ that $L_{\xi_1}\left(\xi_2\iN \theta\right) - L_{\xi_2}\left(\xi_1\iN \theta\right) = \xi_1\wedge \xi_2\iN \omega - d(\xi_1\wedge \xi_2\iN \theta)$. Thus
\begin{equation}\label{calcul2}
\delta \Theta^\Sigma_\Gamma(\delta_1\Gamma,\delta_2\Gamma) = \int_{\Sigma\cap\Gamma}\xi_1\wedge \xi_2\iN \omega - d(\xi_1\wedge \xi_2\iN \theta).
\end{equation}
Hence if we assume that the restriction of $\xi_1$ and $\xi_2$ to $\Sigma_t\cap\Gamma$ has compact support or is rapidly decreasing (as in the preceding paragraph) we obtain (\ref{deltaTheta=Omega}).

Hence we conclude that, under some hypotheses, one can endow the set $\mathcal{E}$ of solutions to the Hamilton equations with a symplectic form $\Omega$ given by(\ref{deltaTheta=Omega}). This form does depend not on $\Sigma$ under the condition that the boundary terms $\int_{\Sigma_t\cap\Gamma} d\left(\xi\wedge \frac{\partial}{\partial t}\iN\theta\right)$ in (\ref{calcul1}) and $-\int_{\Sigma\cap\Gamma}d(\xi_1\wedge \xi_2\iN \theta)$ in (\ref{calcul2}) vanish. This means that, on each slice $\Sigma$, the Jacobi vector fields $\xi,\xi_1,\xi_2$ decreases sufficiently rapidly at infinity. Such a condition is true if, for instance, the manifold $\mathcal{X}$ is a Lorentzien manifold, the slice $\Sigma$ is (a lift of) a spacelike hypersurface of $\mathcal{X}$ and we impose in the definition of $\mathcal{E}$ that all Hamiltonian $n$-curves $\Gamma$ in $\mathcal{E}$ are asymptotic to a given `ground state' Hamiltonian $n$-curve $\Gamma_0$ at infinity on each slice $\Sigma$.

With such a symplectic structure $\Omega$ on $\mathcal{E}$ we can define a Poisson bracket on real-valued functionals on $\mathcal{E}$, which is nothing but (\ref{def-poisson-multi}).

\subsection{A geometric view of the proof}

We can give an alternative proof of Relation (\ref{differencedeTheta}) with a more geometric flavor. We will be even more heuristic, however the validity of our argument is strongly based on the fact that the Lagrangian action can be represented by (\ref{action}). For that purpose imagine that our problem models a hyperbolic time evolution problem and that there are well-defined notions of time and space coordinates on $\mathcal{M}$ (as it is the case for any wave equation on a curved space-time).

\begin{figure}[h]\label{visual}
\begin{center}
\includegraphics[scale=0.5]{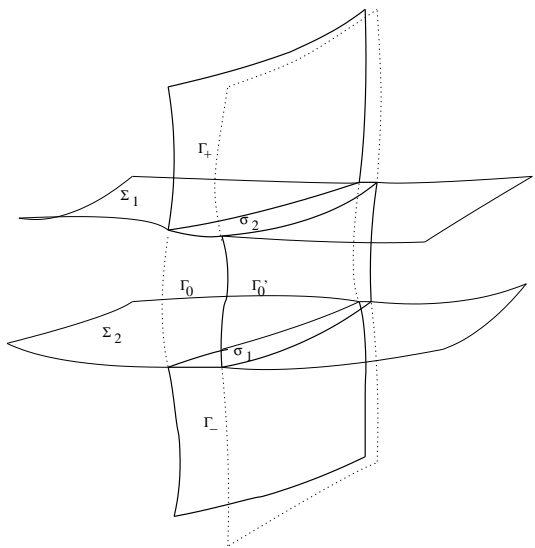}
\caption{\footnotesize A geometric comparison of $\Theta^{\Sigma_1}(\delta\Gamma)$ with $\Theta^{\Sigma_2}(\delta\Gamma)$}
\end{center}
\end{figure}

Consider a Hamiltonian $n$-curve $\Gamma$ and let $\Gamma'$ be another Hamiltonian $n$-curve, which we suppose to be close to $\Gamma$. More precisely we assume that $\Gamma' = \Gamma + \varepsilon \delta\Gamma +o(\varepsilon)$, where $\varepsilon>0$ is a small parameter : by this condition we mean that there exists a vector field $\xi\in \mathcal{C}^\infty(\mathcal{M},T\mathcal{M})$ s.t. $\delta \Gamma = \int_\Gamma\xi$ and $\Gamma'$ is the image of $\Gamma$ by the flow map $e^{\varepsilon\xi}$. We also assume that, for all `time', $\Gamma'$ is asymptotic to $\Gamma$ at infinity in space. Let $\Sigma_1$ and $\Sigma_2$ be two slices, assume that these slices are space-like hypersurfaces and, in order to fix ideas, we suppose that $\Sigma_2$ is in the future of $\Sigma_1$. These slices cross transversally $\Gamma$ and $\Gamma'$ and we denote by $\sigma_1$ (resp. $\sigma_2$) the piece of $\Sigma_1$ (resp. $\Sigma_2$) which is enclosed by the intersections with $\Gamma$ and $\Gamma'$ (see the picture). We also denote by $\Gamma_-$ the part of $\Gamma$ which is in the past of $\Sigma_1$, by $\Gamma_+$ the part of $\Gamma$ which is in the future of $\Sigma_2$ and by $\Gamma_0'$ the part of $\Gamma'$ which is between $\Sigma_1$ and $\Sigma_2$ (see again the picture). Lastly we consider the (not necessarily Hamiltonian) $n$-curve $\Gamma_\varepsilon$, which is the union of $\Gamma_-$, $\sigma_1$, $\Gamma_0'$, $\sigma_2$ and $\Gamma_+$. Of course $\Gamma_\varepsilon$ is not smooth, but it can be approached by a sequence of smooth $n$-curves, so that the following makes sense. We also endow $\Gamma_\varepsilon$ with the orientation which agrees with that of $\Gamma$ on $\Gamma_-\cup\Gamma_+$ and with that $\Gamma'$ on $\Gamma_0'$.

Let us use that fact that $\Gamma$ is a Hamiltonian $n$-curve, hence a critical point of (\ref{action}). It implies that
\begin{equation}\label{visual-1}
\int_{\Gamma_\varepsilon}\theta = \int_\Gamma\theta + o(\varepsilon).
\end{equation}
However the l.h.s. of (\ref{visual-1}) can be decomposed as
\[
\int_{\Gamma_\varepsilon}\theta = \int_{\Gamma_-}\theta + \int_{\sigma_1}\theta + \int_{\Gamma_0'}\theta +\int_{\sigma_2}\theta + \int_{\Gamma_+}\theta,
\]
wheras its r.h.s. is
\[
\int_\Gamma\theta + o(\varepsilon) = \int_{\Gamma_-}\theta +\int_{\Gamma_0}\theta+ \int_{\Gamma_+}\theta + o(\varepsilon),
\]
where $\Gamma_0$ is the part of $\Gamma$ between $\Sigma_1$ and $\Sigma_2$. Hence (\ref{visual-1}) reduces to
\begin{equation}\label{visual-2}
\int_{\sigma_1}\theta + \left(\int_{\Gamma_0'}\theta  - \int_{\Gamma_0}\theta\right) + \int_{\sigma_2}\theta = o(\varepsilon).
\end{equation}
We now recognize that, on the one hand, $\int_{\sigma_1}\theta = \varepsilon \Theta^{\Sigma_1}(\delta\Gamma) +  o(\varepsilon)$, $\int_{\sigma_2}\theta = - \varepsilon \Theta^{\Sigma_2}(\delta\Gamma) +  o(\varepsilon)$ (the sign being due to the orientation of $\sigma_2$). On the other hand $\int_{\Gamma_0}\theta = S_{\Sigma_1}^{\Sigma_2}(\Gamma_0)$ and $\int_{\Gamma_0'}\theta = S_{\Sigma_1}^{\Sigma_2}(\Gamma_0') = S_{\Sigma_1}^{\Sigma_2}(\Gamma_0 + \varepsilon  \delta\Gamma_0) + o(\varepsilon)$. Hence (\ref{visual-2}) gives us
\[
\varepsilon \Theta^{\Sigma_1}(\delta\Gamma) + \varepsilon \left(\delta S_{\Sigma_1}^{\Sigma_2}\right)_\Gamma(\delta\Gamma) - \varepsilon \Theta^{\Sigma_2}(\delta\Gamma) = o(\varepsilon).
\]
Thus by dividing by $\varepsilon$ and letting $\varepsilon$ tend to 0, we recover (\ref{action}).

\section{Geometric quantization}

We address here the question of building a geometric quantization scheme, or at least a prequantization scheme for fields based on the covariant phase space structure. This was more or less the programm envisioned by G. Segal in 1960 \cite{segal}. We present here some attempt of that by using the multisymplectic theory on a very elementary example, which is the one presented in \S \ref{paragraph-example}.\\

\noindent
\textbf{Canonical vector fields on the set of solutions ${\cal E}$}\\

\noindent
We can associate to each $F\in \gP^{n-1}_{\cal
  H}{\cal M}$ a
  tangent vector field $\Xi_F$ on ${\cal E}$ wich is given by
\[
\forall \Gamma\in {\cal E}, \quad
\Xi_F(\Gamma):= \int_\Gamma \xi_F.
\]

\noindent
In the case of the Klein--Gordon equation (\ref{kg}) it is interesting
to represent solutions in ${\cal E}$ by local coordinates. The more
convenient way is based on the Fourier transform: any Hamiltonian
$n$-curve $\Gamma$ is characterized by a solution $\varphi$ to
(\ref{kg}) and by writing
\begin{equation}\label{varphi}
\varphi(x) = {1\over \sqrt{2\pi}^3}\int_{{\cal C}^+} d\mu(k) \left( u_ke^{-ik\cdot x} + u^*_ke^{ik\cdot x}\right),
\end{equation}
we get formally a map
\[
\begin{array}{ccc}
{\cal E} & \longrightarrow & \C^{{\cal C}^+}\times \C^{{\cal C}^+}\\
\varphi & \longmapsto & (u_k,u^*_k)_{k\in {\cal C}^+}.
\end{array}
\]
Note that the image of ${\cal E}$ is characterized by the reality
condition $\overline{u_k} = u^*_k$, $\forall k\in {\cal C}^+$. We can
obviously extend this map to the complexification ${\cal E}^{\C}$ of
${\cal E}$ and then this map is an isomorphism. \\

\noindent
\textbf{The creation and annihilation canonical transformations}\\

\noindent
Now given some function $f\in L^2({\cal C}^+)$ consider
\[
\alpha_f:= \int_{{\cal C}^+} d\mu(k)f(k) \alpha_k
= {i\over \sqrt{2\pi}^3}\int_{{\cal C}^+} d\mu(k)
e^{ik\cdot x} f(k) (p^\mu -i\varphi k^\mu)\beta_\mu.
\]
(Note that the observable functional $a_f$ defined in \S \ref{paragraph-example} is obtained by integration of $\alpha_f$ over a slice.) Then
\[
\xi_f:=\xi_{\alpha_f} = {i\over \sqrt{2\pi}^3}\int_{{\cal C}^+}
d\mu(k) e^{ik\cdot x} f(k)\left( ik^\mu {\partial \over
    \partial p^\mu} - (m^2\varphi + i\eta_{\mu\nu}p^\mu k^\nu){\partial
    \over\partial e} + {\partial \over \partial \varphi}\right)
\]
is completely characterized by the fact that it preserves $\omega$ and
$d{\cal H}$ and through its action on $\varphi$:
\[
d\varphi\left(\xi_f\right) = {i\over \sqrt{2\pi}^3}\int_{{\cal C}^+}
d\mu(k) e^{ik\cdot x} f(k).
\]
We can easily integrate $\xi_f$ on $\mathcal{M}$ and its action on $\mathcal{F}$:
\[
U(s,\varphi) =  \varphi + s {i\over \sqrt{2\pi}^3}\int_{{\cal C}^+}
d\mu(k) e^{ik\cdot x} f(k).
\]
In terms of the coordinates $(u_k,u^*_k)_{k\in {\cal C}^+}$ it gives:
\[
U(s,u_k,u^*_k) = (u_k,u^*_k + isf(k)).
\]
Hence we can symbolically denote
\[
\Xi_f:= \Xi_{\alpha_f} = i\int_{{\cal C}^+} f(k){\partial \over
  \partial u^*_k}. 
\]
There is no integration measure used here, the sign $\int$ stands
uniquely for summing linearly independant vectors: the meaning is that
\[
\Xi_f\left({1\over \sqrt{2\pi}^3}\int_{{\cal C}^+} {d\vec{k}\over
2k^0} \left( u_ke^{-ik\cdot x} + u^*_ke^{ik\cdot x}\right)\right) = {1\over
\sqrt{2\pi}^3}\int_{{\cal C}^+} d\mu(k) if(k) e^{ik\cdot x}.
\]
A completely analogue computation can be done for
\[
\alpha_g^*:= \int_{{\cal C}^+} d\mu(k)g(k) \alpha_k^*
= {-i\over \sqrt{2\pi}^3}\int_{{\cal C}^+} d\mu(k)
e^{-ik\cdot x} g(k) (p^\mu +i\varphi k^\mu)\beta_\mu,
\]
where $g\in L^2({\cal C}^+)$. Denoting $\xi_g^*:=\xi_{\alpha_g^*}$ we
have
\[
d\varphi\left(\xi_g^*\right) = {-i\over \sqrt{2\pi}^3}\int_{{\cal C}^+}
d\mu(k) e^{-ik\cdot x} g(k).
\]
Hence $\Xi_g^*:= \Xi_{\alpha_g^*}$ is given by
\[
\Xi_g^* = -i \int_{{\cal C}^+} g(k){\partial \over \partial u_k}.
\]

\noindent
\textbf{Spacetime translations}\\

\noindent
We now look at the canonical vector fields on ${\cal E}$ associated
with spacetime translations $P_\zeta$, where $\zeta$ is constant vector field on $\mathcal{X}$. We recall that $\xi_{P_\zeta^{(\lambda)}} = \zeta$. We must understand the induced vector field $\Xi_\zeta$ on $\mathcal{F}$. Let $U(s,\cdot )$ be the flow mapping of the vector
field $\zeta$: $U(s,x,\varphi, e,p)
= (s,x + \zeta,\varphi, e,p)$. Then the image of
\[
\Gamma = \{(x,\varphi(x),e(x),p(x))\ |\ x\in {\cal X}\} \subset {\cal
  E}
\]
by $U(x,\cdot)$ is
\[
\Gamma_s = \{(x,\varphi(x-s\zeta), e_s(x),p_s(x))\ |\ x\in {\cal X}\},
\]
where the value of $e_s(x)$ and $p_s(x)$ is completely determined by
the constraint that $\Gamma_s\subset {\cal E}$ and by the knowledge of
$\varphi(x-s\zeta)$. This can be proved by a simple change of
variable. Similarly we determine the action of $\Xi_\zeta$ on
the coordinates $(u_k,u^*_k)_{k\in {\cal C}^+}$ by computing its
action on $\varphi$:
\[
\begin{array}{ccl}
\displaystyle 
\left(\Xi_\zeta\varphi\right)(x) & = & \displaystyle {d\over ds}\left({1\over
    \sqrt{2\pi}^3}\int_{{\cal C}^+} d\mu(k)\left(
    u_ke^{-ik\cdot (x-s\zeta)} + u_k^*e^{ik\cdot
    (x-s\zeta)}\right)\right)_{|s=0}\\
& = & \displaystyle {1\over \sqrt{2\pi}^3} \int_{{\cal C}^+}
d\mu(k)\left(ik\cdot\zeta u_ke^{-ik\cdot (x-s\zeta)} -
  ik\cdot\zeta u_k^*e^{ik\cdot (x-s\zeta)}\right)\\
& = & \displaystyle i\left[\left(\int_{{\cal C}^+}k\cdot\zeta
    \left(u_k{\partial \over \partial u_k} - u^*_k
{\partial \over \partial u_k^*}\right)\right)\varphi\right](x).
\end{array}
\]
Hence
\[
\Xi_\zeta = i\int_{{\cal C}^+}k\cdot\zeta\left(u_k{\partial \over
    \partial u_k} - u^*_k {\partial \over \partial u_k^*}\right).
\]

\noindent
\textbf{Geometric prequantization}\\

\noindent
We recall very briefly the prequantization scheme due to B. Kostant and
J.-M. Souriau (generalizing previous constructions by B.O. Koopman, L. Van Hove and
I. Segal, see \cite{kirillov,snyatycki}). We let $({\cal M},\omega)$ be a simply connected symplectic manifold and we assume for simplicity that there exists a 1-form $\theta$ with $\omega = d\theta$. We consider the trivial bundle ${\cal L}:= {\cal
  M} \times \C$ and denote by $\Gamma({\cal M},{\cal L})$ the set of square
integrable sections of ${\cal L}$. Using $\theta$ we can define a
Hermitian connection $\nabla$ acting on $\Gamma({\cal M},{\cal L})$ by
\[
\forall\xi \in \Gamma({\cal M},T{\cal M}),
\forall \psi\in \Gamma({\cal M},{\cal L}),\quad
\nabla_\xi \psi = \xi\cdot \psi - {i\over \hbar}\theta(\xi) \psi.
\]
Then to each function $F\in {\cal C}^\infty ({\cal M},\R)$ we
associate the operator $\widehat{F}$ acting on $\Gamma({\cal M},{\cal L})$
\[
\widehat{F}\psi = F\psi + {\hbar \over i} \nabla_{\xi_F} \psi
= \left(F - \theta(\xi_F)\right) \psi + {\hbar\over i}\xi_F\cdot \psi,
\]
where $dF + \xi_F\iN \omega = 0$. This construction is called the
prequantization of $({\cal M},\omega)$. For instance if $({\cal M},\omega) =
(\R^{2n}, dp_i\wedge dq^i)$, then $\omega = d\theta$, with $\theta =
p_idq^i$ and $\widehat{q^i} = q^i + i\hbar {\partial \over \partial p_i}$
and $\widehat{p_i} = - i\hbar {\partial \over \partial q^i}$. Of course one
needs further restrictions in order to recover an irreducible representation of the Heisenberg algebra (and hence the standard quantization): this will be the purpose of introducing a polarization and a tensorization of the line bundle $\mathcal{L}$ with the bundle of half volume forms transversal to the leaves of the polarization (see \cite{kirillov,snyatycki}).\\

\noindent
We will propose an extension of this procedure to our setting,
concerned with the quantization of fields. We consider the
trivial bundle ${\cal L}:= {\cal E}^{\C}\times \C$ over ${\cal
  E}^{\C}$, where ${\cal E}^{\C}$ is the complexification of the set
of solutions to the 
Klein--Gordon equation (\ref{kg}) as before. On the set $\Gamma(\mathcal{E}^{\C},\mathcal{L})$ of smooth sections of $\mathcal{L}$ (we are here
relatively vague about the meaning of "smooth") we define a notion of
covariant derivative along any vector field of the type $\Xi_F$, where
$F\in \Gamma({\cal E}^{\C},{\cal L})$ by
\[
\forall \psi\in \Gamma({\cal E}^{\C},{\cal L}), \quad
\nabla_{\Xi_F}\psi:= \Xi_F\cdot \psi - {i\over \hbar}\left(\int_\Sigma
  \xi_F\iN \theta\right) \psi,
\]
where $\theta = \theta_\lambda$. Then we define the prequantization of $F\in \gP^{n-1}_{\cal H}{\cal M}$ to be the operator acting on $\Gamma(\mathcal{E}^\mathbb{C},\mathcal{L})$ by:
\[
\widehat{F}\psi:= \left(\int_\Sigma F\right)\psi + {\hbar \over
  i}\nabla_{\Xi_F}\psi = {\hbar \over i}\Xi_F\cdot \psi + 
\left(\int_\Sigma F - \xi_F\iN \theta\right)\psi.
\]

\noindent
\textbf{Prequantization of the creation and annihilation observables}\\

\noindent
We look here for the expressions of the prequantization of $a_f$ and $a_g^*$ given in \S \ref{paragraph-example}. We first set the fact that if $\varphi$ is given in terms of $(u_k,u^*_k)_{k\in \mathcal{C}^+}$ by (\ref{varphi}), then
\[
{1\over \sqrt{2\pi}^3}\int_{\R^n} e^{-i\vec{k}\cdot
  \vec{x}}\varphi(0,\vec{x}) d\vec{x} = {u_k + u_{\overline{k}}^*\over
  2k^0},
\]
and
\[
{1\over \sqrt{2\pi}^3}\int_{\R^n} e^{-i\vec{k}\cdot
  \vec{x}}p^0(0,\vec{x}) d\vec{x} = {u_k - u_{\overline{k}}^*\over
  2i},
\]
where $\overline{k} = (k^0,-\vec{k})$. We deduce the following
\[
\begin{array}{ccl}
\displaystyle 
\int_{\Sigma\cap \Gamma}\alpha_f & = & \displaystyle {1\over
  \sqrt{2\pi}^3}\int_{\R^n} d\vec{x} \int_{{\cal C}^+} {d\vec{k}\over
  2k^0} e^{-i\vec{k}\cdot \vec{x}}f(k) \left(\varphi(0,\vec{x}) k^0 +
  ip^0(0,\vec{x})\right) \\
% & = &\displaystyle  \int_{{\cal C}^+} {d\vec{k}\over2k^0} f(k){1\over \sqrt{2\pi}^3}  \int_{\R^3} d\vec{x} e^{-i\vec{k}\cdot \vec{x}}\left(\varphi(0,\vec{x}) k^0 + ip^0(0,\vec{x})\right) \\
 & = &\displaystyle  \int_{{\cal C}^+} {d\vec{k}\over2k^0} f(k) \left( {u_k +
  u_{\overline{k}}^*\over 2} + {u_k - u_{\overline{k}}^*\over
  2}\right)\\
 & = & \displaystyle \int_{{\cal C}^+} {d\vec{k}\over2k^0} f(k)u_k.
\end{array}
\]
Similarly
\[
\int_{\Sigma\cap \Gamma}\alpha_g^* = \int_{{\cal C}^+}
{d\vec{k}\over2k^0} g(k)u_k ^*.
\]
We moreover observe that
\[
\xi_f\iN \theta = {1\over \sqrt{2\pi}^3} \int_{{\cal C}^+} {d\vec{k}\over
  2k^0} e^{ik\cdot x}{f(k)\over 2} \left(\varphi k^\mu +
  ip^\mu\right)\beta_\mu  =
  {\alpha_f\over 2},
\]
and similarly $\xi_g^*\iN \theta = {\alpha_g^*\over 2}$. Hence
\[
\int_{\Sigma\cap \Gamma}\xi_f\iN \theta = \int_{{\cal C}^+}
{d\vec{k}\over2k^0} {f(k)\over 2}u_k \quad\hbox{and}\quad
\int_{\Sigma\cap \Gamma}\xi_g^*\iN \theta = \int_{{\cal C}^+}
{d\vec{k}\over2k^0} {g(k)\over 2}u_k ^*.
\]
Using the previous results we can now express, for
$\psi\in \Gamma({\cal E}^{\C},{\cal L})$, 
\[
\nabla_{\Xi_f}\psi:= \Xi_f\cdot \psi - {i\over \hbar}\left(\int_\Sigma
  \xi_f\iN \theta\right) \psi
= i\int_{{\cal C}^+} f(k){\partial \psi \over \partial u_k^*}- {i\over
  \hbar}\left(\int_{{\cal C}^+} {d\vec{k}\over2k^0} {f(k)\over
  2}u_k\right) \psi
\]
and
\[
\nabla_{\Xi_g^*}\psi:= \Xi_g^*\cdot \psi - {i\over \hbar}\left(\int_\Sigma
  \xi_g^*\iN \theta\right) \psi
= -i\int_{{\cal C}^+} g(k){\partial \psi \over \partial u_k}- {i\over
  \hbar}\left(\int_{{\cal C}^+} {d\vec{k}\over2k^0} {g(k)\over
  2}u_k^*\right) \psi.
\]
For the prequantizations we obtain:
\[
\widehat{a}_f\psi = \hbar \int_{{\cal C}^+} f(k){\partial \psi \over \partial
  u_k^*} + \left(\int_{{\cal C}^+} {d\vec{k}\over2k^0} {f(k)\over
  2}u_k\right) \psi,
\]
and
\[
\widehat{a}^*_g\psi = -\hbar \int g(k){\partial \psi \over \partial
  u_k} + \left(\int_{{\cal C}^+} {d\vec{k}\over2k^0} {g(k)\over
  2}u_k^*\right) \psi.
\]
We observe that we have formally $[\widehat{a}_f,\widehat{a}_{f'}] =
[\widehat{a}_g^*,\widehat{a}_{g'}^*] = 0$ and
\[
[\widehat{a}_f,\widehat{a}_g^*] = \hbar \int_{{\cal C}^+}
{d\vec{k}\over2k^0} f(k)g(k).
\]

\noindent
\textbf{Prequantization of the stress-energy tensor}\\

\noindent
It relies on finding the prequantization of $P_\zeta^{(\lambda)} =
\zeta^\mu P_\mu^{(\lambda)}$. In principle one should compute the functionals of
$\int_\Sigma P_\zeta^{(\lambda)}$ and
$\int_{\Sigma}\xi_{P_\zeta^{(\lambda)}}\iN \theta_{\lambda}$. But as 
observed in the previous section we have $P_\zeta^{(\lambda)} = \zeta\iN
\theta_\lambda = \xi_{P_\zeta^{(\lambda)}}\iN \theta_\lambda$ because
$Lie_\zeta\theta_\lambda = 0$. Hence
$\int_{\Sigma} P_\zeta^{(\lambda)} - \xi_{P_\zeta^{(\lambda)}}\iN \theta_\lambda =
0$ and so the prequantization of $P_\zeta^{(\lambda)}$ is just
\[
\widehat{P^{(\lambda)}}_\zeta \psi =  {\hbar\over i}\Xi_\zeta\cdot \psi =
\hbar\int_{{\cal C}^+}k\cdot\zeta\left(u_k{\partial \over
    \partial u_k} - u^*_k {\partial \over \partial u_k^*}\right)\psi.
\]
Note that if we need to compute $\int_{\Sigma}
P^{(\lambda)}_\zeta$, it is more suitable to set $\lambda = 1$, since it gives
then the standard expression for the stress-energy tensor. For
instance, if $\zeta = {\partial \over \partial x^0}$
\[
-\int_{\Sigma\cap \Gamma}P^{(1)}_0 =
 \int_{\R^{n-1}}d\vec{x}\left({(p^0)^2\over 2} + \sum_{i=1}^3
 {(p^i)^2\over 2} + m^2{\varphi^2\over 2}\right)
\]
gives the total energy in the frame associated with the coordinates
$x^\mu$.\\
%The expression for other values of $\lambda$ differs from the integration of an exact form: indeed we have $\theta_1-\theta_\lambda = (1-\lambda) d(p^\nu\varphi\beta_\nu)$ so that
%\[ \begin{array}{ccl}
% P^{(1)}_\mu - P^{(\lambda)}_\mu & = & \displaystyle (1-\lambda){\partial
%  \over \partial x^\mu} 
% \iN d(p^\nu\varphi\beta_\nu)\\
% & = & \displaystyle (1-\lambda)Lie_{\partial \over \partial
%  x^\mu} (p^\nu\varphi\beta_\nu) - (1-\lambda)d\left( {\partial \over \partial
%    x^\mu} \iN p^\nu\varphi\beta_\nu\right)\\
%& = &\displaystyle  -(1-\lambda) d\left( {\partial
% \over \partial x^\mu} \iN p^\nu\varphi\beta_\nu\right).
%\end{array}
% \]

\noindent
\textbf{Introducing a polarization}\\

\noindent
We choose to impose the extra condition $\nabla_{\Xi_g^*}\psi = 0$, $\forall g$ (covariantly antiholomorphic sections), which gives us:
\[
\psi(u_k,u_k^*) = h(u_k^*)\exp\left({ -{1\over 2\hbar}\int_{{\cal C}^+}
  {d\vec{k}\over2k^0} u_ku_k^*}\right) = h(u_k^*)|0\rangle.
\]
The advantage of this choice is that all observables functional (creation and annihilation, energy and momentum) of are at most linear in the variables $(u_k,u_k^*)$, hence we do not need to use the Blattner--Kostant--Sternberg correction for these operators \cite{kirillov,snyatycki}. As a result $\widehat{P}|0\rangle = 0$, so that the energy of the vacuum vanishes without requiring normal ordering. However we did not take into account the metaplectic correction, which requires a slight change of the connection: if we would do that we would find that the vacuum as an infinite energy (as in the standard quantization scheme), which can be removed by a normal ordering procedure. The mysterious thing here (as was already observed) is that by ignoring the metaplectic correction (which however is fundamental for many reasons) we do not need the normal ordering correction.

\end{document}